\pdfoutput=1 
\documentclass[11pt]{article}
\usepackage[utf8]{inputenc}
\usepackage{typearea}
\usepackage{amsmath,amsfonts,amssymb,amscd}
\usepackage[all]{xy}
\usepackage{graphicx}
\usepackage{cite}
\usepackage{mathrsfs}
\usepackage{bbm}
\usepackage{bbold}
\usepackage{braket}
\usepackage[dvipsnames]{xcolor}

\usepackage[normalem]{ulem}
\usepackage{tensor}
\usepackage{mathtools}
\usepackage[bookmarks=false]{hyperref}
\usepackage{srcltx}

\newcommand{\be}{\begin{equation}}
\newcommand{\ee}{\end{equation}}
\newcommand{\bea}{\begin{eqnarray}}
\newcommand{\eea}{\end{eqnarray}}

\newcommand{\E}{\mathcal{E}}
\newcommand{\F}{\mathcal{F}}
\newcommand{\B}{\mathcal{B}}
\renewcommand{\H}{\mathcal{H}}
\newcommand{\C}{\mathbb{C}}
\newcommand{\n}{\mathbf{n}}
\newcommand{\m}{\mathbf{m}}

\DeclareMathOperator{\Tr}{Tr}

\renewcommand{\S}{\mathcal{S}}

\newcommand{\ketbra}{\rangle \langle}

\newcommand{\choiE}{\widetilde {\cal E}}
\newcommand{\choiI}{\widetilde {\cal I}}

\newcommand{\ini}{\text{in}}
\newcommand{\out}{\text{out}}
\newcommand{\Hin}{\mathcal{H}_\ini}
\newcommand{\Hout}{\mathcal{H}_\out}
\newcommand{\rin}{\rho_\ini}
\newcommand{\rout}{\rho_\out}
\newcommand{\riSE}{\overline{\rho}_\ini}
\newcommand{\roSE}{\overline{\rho}_\out}
\newcommand{\din}{d_\ini}
\newcommand{\dout}{d_\out}
\newcommand{\pin}{p_\ini}
\newcommand{\tildepin}{\tilde{p}_\ini}
\newcommand{\tildepout}{\roSE}

\definecolor{dg}{rgb}{0,.5,0}

\usepackage{soul}

\def\mytitle{Prospects for quantum process tomography \\[2mm] 
at high energies}

\title{\mytitle}

\begin{document}

\begin{titlepage}
\setlength{\topmargin}{0.0 true in}
\thispagestyle{empty}

\begin{center}
{\LARGE\textbf{\mytitle}}

\renewcommand*{\thefootnote}{\fnsymbol{footnote}}

\vspace{1.1cm}
\large
Clelia Altomonte${}^{1}$\footnote[1]{
\href{clelia.1.altomonte@kcl.ac.uk}{clelia.1.altomonte@kcl.ac.uk}
},~
Alan J.\ Barr${}^{2,}$\footnote[2]{
\href{alan.barr@physics.ox.ac.uk}{alan.barr@physics.ox.ac.uk}
},~
Micha{\l} Eckstein${}^{3,}$\footnote[2]{
\href{michal@eckstein.pl}{michal@eckstein.pl}
}, \\[3mm]
Pawe{\l} Horodecki${}^{4,5}$\footnote[3]{
\href{pawhorod@pg.edu.pl}{pawhorod@pg.edu.pl}
},~and~
Kazuki Sakurai${}^{6,}$\footnote[4]{
\href{kazuki.sakurai@fuw.edu.pl}{kazuki.sakurai@fuw.edu.pl}
}
\\[5mm]

\normalsize

\textit{
${}^1$Department of Physics, Strand, King’s College London, WC2R 2LS
} \\[2mm]
\textit{
${}^2$Department of Physics, Keble Road, University of Oxford, OX1 3RH
\\
Merton College, Merton Street, Oxford, OX1 4JD
} \\[2mm]
\textit{
${}^3$
Institute of Theoretical Physics, Faculty of Physics, Astronomy and Applied Computer Science, 
Jagiellonian University, ul.\ {\L}ojasiewicza 11, 30–348 Kraków, Poland
} \\[2mm]
\textit{
${}^4$
International  Centre  for  Theory  of  Quantum  Technologies,  University  of  Gda\'nsk,  Wita  Stwosza  63,  80-308  Gda\'nsk,  Poland
} \\[2mm]
\textit{
${}^5$
Faculty of Applied Physics and Mathematics, National Quantum Information Centre,
Gda\'nsk University of Technology, Gabriela Narutowicza 11/12, 80-233 Gda\'nsk, Poland
} \\[2mm]
\textit{
${}^6$Institute of Theoretical Physics, Faculty of Physics,\\
University of Warsaw, Pasteura 5, 02-093, Warsaw, Poland
}

\end{center}
\vspace*{5mm}

\begin{abstract}\noindent

In quantum information theory, the evolution of an open quantum system -- a unitary evolution followed by a measurement -- is described by a quantum channel or, more generally, a quantum instrument. 
In this work, we formulate spin and flavour measurements in collider experiments as  quantum instruments.   
We demonstrate that the Choi matrix, which completely determines input-output transitions, can be both theoretically computed from a given model and experimentally reconstructed from a set of final state measurements (quantum state tomography) using varied input states.
The experimental reconstruction of the Choi matrix, known as quantum process tomography, offers a powerful new approach for probing potential extensions of the Standard Model within the quantum field theory framework and at the same time constitutes a new foundational test of quantum mechanics itself. 
As an example, we outline the quantum process tomography approach applied to the $e^+ e^- \to t \bar{t}$ process at a polarized lepton collider.

\end{abstract}

\thispagestyle{empty}
\clearpage

\end{titlepage}

\section{Introduction and motivation}

Particle accelerators and colliders have revolutionised our understanding of the microscopic world, helping us understand nature at the smallest scales. The high-energy interactions that are probed with these machines provide the most direct evidence of the most basic quantum fields and their interactions. 

While it is well understood that quantum mechanical processes dominate on these scales, and the calculations of them are based in the language of quantum field theory, it has only been recently, with the advent of advances in quantum computational technologies, that the potential of treating colliders as natural quantum information processors is starting to be investigated in detail. 

There has recently been a surge of interest~\cite{Afik:2020onf, Ashby-Pickering:2022umy, Barr:2021zcp, Barr:2022wyq, Afik:2022kwm, Aguilar-Saavedra:2022uye, Aoude:2022imd, Severi:2022qjy, Fabbrichesi:2022ovb, Altakach:2022ywa, Sakurai:2023nsc, Aoude:2023hxv, Bernal:2023ruk, Han:2023fci, Dong:2023xiw, Fabbrichesi:2023jep, Bi:2023uop, Maltoni:2024tul, Afik:2024uif, Maltoni:2024csn, Morales:2024jhj, Aguilar-Saavedra:2024fig, Aguilar-Saavedra:2024hwd, Aguilar-Saavedra:2024vpd, Aguilar-Saavedra:2024whi, Grabarczyk:2024wnk} in making explicitly quantum-sensitive measurements at colliders, including measurements of entanglement and Bell-inequalities (for a recent review see Ref.~\cite{Barr:2024djo}). New experimental results, such as the observation of top-quark entanglement at the ATLAS and CMS experiments at the Large Hadron Collider~\cite{ATLAS:2023fsd,CMS:2024pts,CMS:2024zkc} have demonstrated the potential of making measurements of explicitly quantum phenomena at the highest energies.

This new set of proposed (and realised) measurements generally investigates quantum features, such as spin or flavour, which span a low-dimensional Hilbert space. 
Many of the recent proposals are based on the observation that a weakly-decaying particle, when it decays through a chiral coupling, offers a natural mechanism for measuring the spin of the parent. 
Thus by measuring the momentum of the daughter particles of a large ensemble of decays -- an ensemble that is typically available at colliders --  one can reconstruct the full spin density matrix of the parent, the processes known as `quantum state tomography' in the quantum information literature. By performing correlated measurements on multiple such weakly-decaying parents, it is possible to reconstruct the full spin-density matrix for a system of several particles and to interrogate their quantum properties.

To treat the collider as a quantum computer, one must go beyond measurements of those particles produced in the collision and investigate the physical process that created those particles, involving state preparation, unitary evolution, and then measurement~\cite{AltomonteBarr23}. 
To maximally exploit this correspondence, one needs to extend the experimentalist's role beyond the use of colliders as ``observatories'' of collisions, in which they perform a passive role of observing the results of collisions. 
One then wishes to determine the quantum properties of the output for different input states -- a procedure known in quantum information theory as \textit{quantum process tomography}~\cite{PTomography1,PTomography2}.

Foundational tests of the quantum theory, such as the Bell test, demand that experiments have tunable parameters that can be modified by the experimenter either at the time of state preparation, evolution or measurement (see e.g.\ \cite{AspectBell2015}). It is clearly more challenging for the experimentalist to modify such parameters in high-energy experiments than in desktop experiments. Even so, there is motivation to make steps in this direction since it would open the door to performing direct tests of the quantum-mechanical paradigm in high energy physics. 
We will present one such foundational test that involves initial state preparation by polarising the spins of the incoming particle beams~\cite{Eckstein:2021pgm}.

There are good reasons to want to investigate the quantum computational properties of colliders. 
Most obviously, we can use these methods to test the Standard Model of particle physics. 
For many quantum processes one expects to be able to make measurements and compare them to high-precision theoretical predictions made using perturbation theory. Any deviation would provide evidence for new fields or phenomena present in nature but not yet included in the Standard Model. It is worth noting that even when the energy of the collider is not sufficient to excite new resonances of any such fields, they are still expected to provide loop contributions to scattering amplitudes. Performing tests of this nature already forms a major area of research at colliders. Using techniques from quantum information theory it has been shown to sharpen sensitivity to some such models --- a variety of examples may be found in Section~7 of Ref~\cite{Barr:2024djo}.

Second, one wishes to be able to test for deviations from predictions not just of a particular quantum field theory (e.g.\ the Standard Model) but of quantum theory in general. There are many fundamental predictions of quantum mechanics, such as entanglement, linearity, complete positivity, and the Born rule for measurement statistics, that can be directly tested by treating colliders as quantum-information processing devices \cite{Eckstein:2021pgm}. A deviation from any of these predictions might point towards new physics based on a `beyond-quantum' theory, such as a general probabilistic theory \cite{GPT_review}.  In contrast to most other tests in high-energy physics, these types of tests do not require that the channel dynamics is calculable since they are tests of the validity of the concept of a quantum channel itself. 
As a result, they can be performed for a wider range of processes, such as in low-energy hadron physics, and are not constrained to be calculable in perturbation theory.

Finally, by working to bridge techniques in modern collider physics and quantum information theory, we may be able to provide improved insights into each of those domains.

The paper is organised as follows. Section~\ref{sec:review} reviews quantum channels, quantum instruments and Choi matrices, while Section~\ref{sec:QIcol} describes the application of those concepts to the collider environment. An example procedure for quantum process tomography in $e^+e^- \rightarrow t\bar{t}$ events is described in Section~\ref{sec:tomography}, along with the predictions for that process from perturbation theory within the Standard Model. Some practical considerations for undertaking experiments with polarised beams are discussed in Section~\ref{sec:practicalities}. 

\section{A review on quantum maps, Choi matrices and process tomography}
\label{sec:review}

\subsection{Quantum channels} 

In quantum information theory 
 (see e.g.\ \cite{nielsen00,qubitguide}), the general evolution of a subsystem, which may be interacting with another system (e.g.\ environment), is described by the {\it quantum channel}.
Let ${\cal S}({\cal H})$ be the set of density operators (i.e.\ quantum states) on a Hilbert space, $\cal H$.
The quantum channel, ${\cal E}$, is a linear map from an initial density operator to a final one, ${\cal E}: {\cal S}(\Hin) \to {\cal S}(\Hout)$.
As indicated, the Hilbert spaces for the initial and final states need not be the same. 
Quantum channels can be obtained by the following operational steps:
\begin{description}

\item[ {[0]} ]

We prepare the system in the initial state, $\rin \in {\cal S}(\Hin)$.

\item[ {[1]} ]  

We embed the state in a total Hilbert space ${\cal H}_E \otimes \Hin$ involving an auxiliary system --- `environment' --- and assume that initially there are no correlations between the prepared system and its environment\footnote{This is a standard and important assumption underlying the idea of a quantum channel \cite{PTomography1}.}.  
By taking a large enough auxiliary system, one can always assume the state of the auxiliary system to be pure, denoted by $\ket{0}_E$, so that
$$
\rin \,\mapsto\, \riSE \,=\, | 0 \ketbra 0|_E \otimes \rin .
$$

\item[ {[2]} ]  
We assume the total system to be closed so that $\overline{\rho}$ evolves with a unitary transformation on ${\cal H}_E \otimes \Hin$,
$$
\riSE \,\mapsto\, \roSE \,=\, U  \riSE U^\dagger,
$$
with $U^\dagger U = U U^\dagger = {\mathbf 1}$.

\item[ {[3']} ] 
Finally, we trace out that part of the total Hilbert space which we are not interested in, or do not have access to, to obtain:
$$
\roSE \,\mapsto\, \rout \,=\, {\rm Tr}_{E'}  (\roSE)
\,\in\, {\cal S}(\Hout) \,,
$$
with ${\cal H}_{E'} \otimes \Hout = {\cal H}_{E} \otimes \Hin$.

\end{description}

This procedure indeed defines a map, ${\cal E}(\rin) = \rout$, for any $\rin \in \mathcal{S}(\Hin)$.
To compute ${\Tr}_{E'}$ in the last step explicitly, 
one can introduce a basis of ${\cal H}_{E'}$,  $\{ \ket{k} \}$.
We have
\be
{\Tr}_{E'}(\roSE)
\,=\,
\sum_k  \bra{k} U \left( | 0 \ketbra 0|_E \otimes \rin \right) U^\dagger \ket{k} \,.
\ee
Introducing operators $E_k = \bra{k} U \ket{0}_E$,
one can therefore write
\be
\rout \,=\, \sum_k E_k \rin E_k^\dagger     \,.
\ee
The operators $E_k$ are called Kraus operators, which map states from $\Hin$ to those in $\Hout$.
One can easily see that they satisfy the normalisation 
$\sum_k E_k^\dagger E_k = {\bf 1}$.
This expression makes it clear that the map is trace-preserving, 
$\Tr \rout = \Tr ( \sum_k E_k^\dagger E_k \rin ) = \Tr (\rin )$,
and positive, $\rout = \sum_k (E_k \sqrt{\rin}) (E_k \sqrt{\rin})^\dagger$.
In fact, it is {\it completely positive}, meaning that the positivity persists even if the map acts on a part of a larger system: 
For $\tilde\rho  _\ini = \rho_{\rm ext} \otimes \rin$,
the map is extended with the Kraus operators, $\tilde E_k \equiv {\bf 1}_{\rm ext} \otimes E_k$, as
$\tilde\rho  _\ini\mapsto \tilde\rho  _\out = \sum_k \tilde E_k \tilde\rho  _\ini \tilde E_k^\dagger$, and one can see that $\tilde\rho  _\out$ is positive as before.

\subsection{Quantum instruments} 

{\it Quantum instruments} \cite{Davies76}, also known as {\it quantum operations} \cite{PTomography1}, are similar to quantum channels, but they differ in the last step {\bf [3']} of the above procedure.
Instead of tracing out the auxiliary system $E'$, we make a {\it selective} measurement on it.
Let ${\cal P}_x$ be some projection operator acting on ${\cal H}_{E'}$ with some range, $x$, of a classical outcome. If $x$ is discrete, then one could write ${\cal P}_x = \sum_{k \in x}| k \ketbra k |$ with normalised eigenstates $\ket{k} \in {\cal H}_{E'}$.
The step {\bf [3']} of the above procedure is then replaced by
\begin{description}

\item[ {[3]} ] 
We make a selective measurement on $E'$, requiring that the measurement outcome lies in $x$ and trace out the Hilbert space ${\cal H}(E')$:
$$
\roSE \,\mapsto\, \varrho_x \,\equiv\, {\rm Tr}_{E'} \left[ {\cal P}_x \roSE {\cal P}_x \right] \,.
$$
\end{description}
This map is linear but, in general, not trace-preserving. 
In fact, quantum instruments are characterised as completely positive trace non-increasing maps: $0 < \Tr \varrho_x \leq \Tr \rin$. 
Hence, the quantum instrument is a map ${\cal I}_x: {\cal S}(\Hin) \to {\cal B}_+(\Hout)$, where ${\cal B}_+({\cal H})$ is the set of positive bounded operators on ${\cal H}$.

The post-measurement state, $\rho_x$, is obtained by renormalising $\varrho_x$: 
\be
\rho_x \,=\, \frac{ \varrho_x }{ \Tr \varrho_x } ,
\label{poststate}
\ee
However, the map $\rin \mapsto \rho_x$ is not linear due to the presence of $\rin$ in the denominator, $\Tr \varrho_x = \Tr ( {\cal P}_x U (|0 \ketbra 0 |_E \otimes \rin) U^\dagger )$. Note that the latter quantity is simply the probability of registering the classical outcome within the range $x$, given that the initial state was $\rin$.

A quantum instrument is related to a quantum channel in an obvious way: 
If we were to include the outcomes $\overline{x}$ we have discarded, ${\cal P}_{\bar x} = {\bf 1} - {\cal P}_{x} = \sum_{k \in \bar x} | k \ketbra k|$, the map would become a quantum channel:
$\roSE \mapsto \rout = \Tr_{E'}(( {\cal P}_x + {\cal P}_{\bar x}  ) \roSE ) $.
The quantum instrument can also be constructed with the restricted set of Kraus operators corresponding to the domain~$x$:
\be
\varrho_x \,=\, {\cal I}_x(\rin) \,=\, \sum_{k \in x} E_k \rin E_k^\dagger\,.
\label{Unnormalised-State}
\ee

If the range of outcomes $x$ that we are interested in is a continuous set, as it will be the case in Section \ref{sec:QIcol}, then the sums over $k \in x$ should be replaced by integrals $\int_x d \mu(k)$ with an appropriate measure $\mu$.

\subsection{Choi matrices} 

Let $\din$ and $\dout$ be the dimensions of the input and output Hilbert spaces: $\din = \dim \Hin$ and $\dout = \dim \Hout$.
Any linear map on operators ${\cal B}(\Hin) \to {\cal B}(\Hout)$ can be described, in chosen bases of $\Hin$ and $\Hout$, through a $D \times D$ matrix with $D = \din \cdot \dout$.
For the case of a completely positive map, such a matrix is called a {\it Choi matrix} of the map.

Let $\{\ket{i}\}$ 
and $\{\ket{\alpha}\}$ 
be the orthonormal bases of $\Hin$ and $\Hout$, respectively. 
We first expand the input state in the basis $\{|i \ketbra j|\}$  of ${\cal B}(\Hin)$: $\rin = \sum_{i,j} \rho_{ij} | i \ketbra j| $,
where $[\rho_{ij}]$ is a $\din \times \din$ density matrix. 
The quantum channel, $\cal E$, is completely determined once its action on all the basis elements, ${\cal E}(|i \ketbra j|)$, is specified. Then, by linearity, we calculate the action of $\cal E$ on any state $\rin$ as
\be
{\cal E}( \rin ) = \sum_{i,j} \rho_{ij} \, {\cal E}(|i \ketbra j|) 
\,.
\ee
Since ${\cal E}(|i \ketbra j|) \in {\cal B}(\Hout)$ it can be represented by a $\dout \times \dout$ matrix, $[{\cal E}(|i \ketbra j|)]_{\alpha \beta} = \bra{\alpha} {\cal E}(|i \ketbra j|) \ket{\beta}$.
The $D \times D$ Choi matrix, $\choiE$, is then expressed in this basis as
\be \label{ChoiE}
\choiE \,\equiv\, \frac{1}{\din} \sum_{i,j} | i \ketbra j |  \otimes {\cal E}(| i \ketbra j |) = 
\frac{1}{\din}
\begin{pmatrix}
{\cal E}(|1 \ketbra 1|) & {\cal E}(|1 \ketbra 2|)  & \cdots & {\cal E}(|1 \ketbra \din|) \\
{\cal E}(|2 \ketbra 1|) & {\cal E}(|2 \ketbra 2|) & \cdots & {\cal E}(|2 \ketbra \din|)  \\
\vdots & \vdots & \ddots & \vdots\\
{\cal E}(|\din \ketbra 1|) & {\cal E}(|\din \ketbra 2|) & \cdots & {\cal E}(|\din \ketbra \din|)  \\
\end{pmatrix}.
\ee
Here, each element ${\cal E}(|i \ketbra j|)$ is a $\dout \times \dout$ block matrix. 
Since ${\cal E}$ is trace-preserving, the trace of the diagonal block is one, $\Tr [{\cal E}(|i \ketbra i|)] = 1$, for any $i \in \{1,\ldots,\din\}$,
which means that the Choi matrix has a unit trace. 


A powerful theorem in quantum information, known as the Choi--Jamio{\l}kowski isomorphism \cite{JAMIOLKOWSKI1972275,CHOI1975285} or {\it channel--state duality}, states that a map $\E: {\cal B}(\Hin) \to {\cal B}(\Hout)$ is completely positive if and only if the corresponding Choi matrix \eqref{ChoiE} is positive semi-definite. In particular, $\E$ is a quantum channel iff $\choiE$ is a quantum state in $S(\Hin \otimes \Hout)$.
%
Consequently, it can also be written as a mixture of pure states $\ket{\psi_k} \in \Hin \otimes \Hout$ with some probabilities $p_k$,
\be
\choiE = \sum_k | \tilde \psi_k \ketbra \tilde \psi_k |, 
\label{choi_mix}
\ee
where $\ket{\tilde \psi_k} \equiv \sqrt{p_k} \ket{\psi_k}$.

We note that any ket $\ket{\tilde \psi_k} \in \Hin \otimes \Hout$ can be obtained from the maximally entangled state in $\Hin \otimes \Hin$,
\be
\ket{\Omega} \,\equiv\, \frac{1}{\sqrt{\din}} \sum_{i=1}^{\din} \ket{i} \otimes \ket{i} 
\ee
through the action of some operator $E_k: \Hin \to \Hout$ as 
\be
\ket{\tilde \psi_k} \,=\, ({\bf 1} \otimes E_k) \ket{ \Omega }.
\label{EkO} 
\ee
Using this expression, Eq.\ \eqref{choi_mix} becomes
\bea
\choiE =  \sum_k ({\bf 1} \otimes E_k) | \Omega \ketbra \Omega | ({\bf 1} \otimes E_k^\dagger)
\nonumber 
=
\frac{1}{\din} \sum_{i,j} | i \ketbra j | \otimes \left( \sum_k E_k | i \ketbra j | E_k^\dagger \right)
\eea
Comparing the last expression with Eq.\ \eqref{ChoiE}, we conclude that $E_k$ are the Kraus operators for the channel $\cal E$.
In fact, $\Tr \choiE = 1$ implies $\sum_k E_k^\dagger E_k = {\bf 1}$.

Similarly, one can construct the $D \times D$ Choi matrix for a quantum instrument
\be\label{ChoiI}
\widetilde {\cal I}_x \,\equiv\,
\frac{1}{\din}
\begin{pmatrix}
{\cal I}_x(|1 \ketbra 1|) & {\cal I}_x(|1 \ketbra 2|)  & \cdots & {\cal I}_x(|1 \ketbra \din|) \\
{\cal I}_x(|2 \ketbra 1|) & {\cal I}_x(|2 \ketbra 2|) & \cdots & {\cal I}_x(|2 \ketbra \din|)  \\
\vdots & \vdots & \ddots & \vdots\\
{\cal I}_x(|\din \ketbra 1|) & {\cal I}_x(|\din \ketbra 2|) & \cdots & {\cal I}_x(|\din \ketbra \din|)  \\
\end{pmatrix}.
\ee

The Choi--Jamio{\l}kowski isomorphism \cite{JAMIOLKOWSKI1972275,CHOI1975285} applied to quantum instruments implies that the map ${\cal I}_x: {\cal S}(\Hin) \to {\cal B}_+(\Hout)$ is completely positive if and only if the corresponding Choi matrix \eqref{ChoiI} is positive semi-definite.

%
%

\subsection{Quantum process tomography}\label{sec:tomo}

Suppose now that we have an unknown quantum-information processing device (called a ``Q-data box'' in Ref.~\cite{Eckstein:2021pgm}) at hand, which we can probe with a programmable quantum system. The device outputs another quantum system,  which can be studied through quantum projective measurements. The effective Hilbert spaces at the input and output are $\Hin$ and $\Hout$, respectively. If the device implements some quantum channel $\E: S(\Hin) \to S(\Hout)$, then it is completely determined by the $D \times D$ matrix \eqref{ChoiE}. 
In fact, because of the positivity and unit trace of $\widetilde{\E}$, the number of independent real parameters needed to characterise $\E$ is $\din^2 (\dout^2-1)$, see~\cite{PTomography1}.

The goal of quantum process tomography \cite{PTomography1,PTomography2} is to reconstruct (or ``estimate'') the quantum channel $\E$ by experimentally measuring the relevant parameters. The basic step of the procedure is as follows:

\begin{enumerate}

	\item Prepare the probe system in a quantum state $\rin^k \in S(\Hin)$.
	
	\item Pass it through the device (Q-data box).
	
	\item Measure the observable corresponding to 
              $M_n \in \B_+(\Hout)$ on the outgoing system.

\end{enumerate}

Suppose now that we prepare a large number of probe systems in the same initial state $\rin^1$. 
For every such input we make measurements from a suitable set $\{M_n\}_{n=1}^{\dout^2-1}$, which is ``tomographically complete'', that is, it constitutes a basis of the space $\S(\Hout)$.
To better estimate the corresponding Choi matrix elements, we must perform these measurements multiple times on an ensemble of output states for each input state.
From the gathered statistics we can reconstruct the {\it effective} state of the outgoing system, expressed as a density matrix in the basis determined by $M_n$'s. This is known as the {\it quantum state tomography} or {\it quantum state estimation} procedure \cite{Tomo2004}.  
In the context of the states of decaying massive particles, one cannot choose the axis of decay -- which also corresponds to the axis of measurement -- and so the quantum state tomography is obtained instead from the outcomes of a set of measurements in many different directions, with each corresponding to the momentum direction of an emitted particle~\cite{Afik:2020onf,Ashby-Pickering:2022umy}.

Once we are satisfied with the estimate $\rout^1$ of the final state for the initial state $\rin^1$, we repeat the procedure with another initial state, $\rin^2$. 

In order to reconstruct the quantum channel $\E$ we need $\din^2$ linearly independent initial states $\{\rin^k\}_{k=1}^{\din^2}$, which form a basis of the space $\B(\Hin)$. 
At the end of the day, we are left with a dataset consisting of $\din^2$ pairs of input and output quantum states, $\{(\rin^k,\rout^k)\}_{k=1}^{\din^2}$. If $\E$ is indeed a quantum channel, then we have $\rout^k = \E(\rin^k)$, for every~$k$.

Because $\rin^k$'s form a basis of ${\cal B}({\cal H_{}})$ we can write
\be
\quad |i \ketbra j| = \sum_{k=1}^{\din^2} X^{(i,j)}_k \, \rin^k,
\quad \text{ with some coefficients } \quad X^{(i,j)}_k \in \mathbb{C}. \label{TheNewX}
\ee
Then, by the assumed linearity of $\E$, the output of the channel is completely determined by its action on the basis states 
\be
\E( |i \ketbra j| ) = \sum_{k=1}^{\din^2} X^{(i,j)}_k \, \rout^k.
\label{TheNewX2}
\ee
In Appendix \ref{app:tomo} we present the complete method to reconstruct the corresponding Choi matrix $\widetilde{\E}$ from an arbitrary set of mixed initial states, which span the space $S(\Hin)$. 

Along the same lines, one can perform the quantum process tomography of a quantum instrument. 
In this case, however, one needs an additional parameter for every initial $\rin^k$ to determine the trace of each block of the Choi matrix \eqref{ChoiI}.
More precisely, for every prepared initial state $\rin^k$, and a fixed $x$, the quantum state tomography will give us the corresponding effective output state $\rho_x^k$. In order to 
obtain $\varrho_x^k = \mathcal{I}_x(\rin^k)$ instead, we need to multiply $\rho_x^k$ by the probability that measurement on the auxiliary system $E'$ gave an outcome in $x$, as indicated in Eq.\ \eqref{poststate}.

In Section \ref{sec:tomography} we present an explicit procedure for the reconstruction of a quantum instrument associated with the $t\bar{t}$ production in polarised $e^-e^+$ scattering. 
In particular, we will see that, in this case, the presence of the outcome $x$ corresponds to a particular energy-momentum selection of the $t\bar{t}$ state.

\section{Quantum instruments in collisions}
\label{sec:QIcol}

\subsection{From $S$-matrix to quantum instruments}\label{sec:S}

In this section, we demonstrate how quantum instruments naturally arise in measurements in high energy experiments. 
For concreteness, we consider $2 \to 2$ scattering\footnote{Extensions to $2 \to n$ scattering and to inclusive processes, e.g.\ 
$\alpha \beta \to \delta \gamma + X$ with $X$ being extra particles to be ignored are straightforward by modifying the projection operator \eqref{proj} as desired.} 
$\alpha \beta \to \gamma \delta$. 
We concentrate on the transition of internal (finite-dimensional) degrees of freedom, such as spin and flavour. 
For the initial state, let us denote the Hilbert space of these internal degrees of freedom by $\Hin = {\cal H}_{\alpha} \otimes {\cal H}_{\beta}$,
and its basis kets by
$\ket{I,J} = \ket{I} \otimes \ket{J}$
or $\ket{K,L} = \ket{K} \otimes \ket{L}$, where $\ket{I}, \ket{K} \in {\cal H}_{\alpha}$ and $\ket{J}, \ket{L} \in {\cal H}_{\beta}$.
Similarly, for the final state, we have
$\Hout = {\cal H}_{\gamma} \otimes {\cal H}_{\delta}$
and 
$\ket{A,B} = \ket{A} \otimes \ket{B}$, $\ket{C,D} = \ket{C} \otimes \ket{D}$
with $\ket{A},\ket{C} \in {\cal H}_{\gamma}$ and $\ket{B},\ket{D} \in {\cal H}_{\delta}$.
We are interested in the quantum instrument, which maps an initial state $\rin \in {\cal S}(\Hin)$
into an unnormalised state $\varrho_x \in {\cal B}_+(\Hout)$,
arising from the time evolution and the subsequent selective measurement 
%
with the outcome lying within a range $x$.



To explicitly construct a quantum instrument, we expand the initial state in terms of the basis operators in $\Hin$ as
\be
\rin = \sum_{_{I,J,K,L}} {\rin}_{[I,J],[K,L]} | I,J \ketbra K,L |,
\label{expand}
\ee
where ${\rin}_{[I,J],[K,L]}$ is the expansion coefficient, also interpreted as the ${[I,J],[K,L]}$ element of the density matrix. 
In what follows, we assume the momentum and the internal degrees of freedom in the initial state are not correlated.
This condition is often satisfied to a good approximation at colliders.
At lepton colliders, the beam energy resolution is good enough so that we can regard the initial state as a pure state, $\ket{\tildepin}$, where $\tildepin$ collectively denotes the momenta of two colliding particles, $\alpha$ and $\beta$. 
The tilde indicates that this state is normalised as
$\braket{ \tildepin | \tildepin } = 1$.
 In the context of high energy collisions, the initial state is of the product form, $\rin = \rin^\alpha \otimes \rin^\beta$, with $\rin^\alpha \in {\cal S}({\cal H}_\alpha)$ and $\rin^\beta \in {\cal S}({\cal H}_\beta)$, because the beams are not correlated before the collision.

In collider physics, the process of calculating a final state $\varrho_x$ within quantum theory proceeds as follows:
\begin{description}
\item[ {[1]} ] 
The initial state of the incoming particles is written as
\be
\riSE \,=\, | \tildepin \ketbra \tildepin | \otimes \rin
\,=\,
 \sum_{_{I,J,K,L}} {\rin}_{[I,J],[K,L]} | \tildepin ; I,J \ketbra \tildepin ; K,L | \,,
\ee
which includes both their momenta and spins.
The condition $\Tr \riSE = 1$ follows from the proper normalisation of $\ket{\tildepin}$.

\item[ {[2]} ] 
We then evolve the entire system with the unitary $S$-matrix:
\be
\riSE \,\mapsto\, \roSE
= S  \riSE  S^\dagger.
\label{preout}
\ee

\item[ {[3]} ] 
After the collision of $\alpha \beta$, we measure the $\gamma \delta$ final state. 
At this point, we might want to constrain the final state momenta in a restricted region $x$. 
This may reflect the fact that the detector has a finite resolution or a fiducial volume, or it is simply due to the event selection in the analysis.  
The complete set in terms of the late-time asymptotic states is given by
\be
\hat {\bf 1} = \sum_{f} 
\left[ 
\left(
\prod_{i \in f} \int  
d \Pi_i
\right)
| f \ketbra f | \right]\,,~~~~
d \Pi_i = \frac{d^3 {\bf p}_i}{(2 \pi)^3 2 E_i} ,  
\label{compset}
\ee
where $\sum_f$ represents the summation over all single- and multi-particle final states and all internal degrees of freedom.
The index $i$ labels an individual particle in the final state $f$.
The $N$ and $N'$ particle states $\ket{f}$ and $\ket{f'}$ have no overlap if their particle contents are different, otherwise  
\be
\braket{f|f'} 
=
\prod_{i \in f} (2 \pi)^3 2 E_i \delta^3( {\bf p}_i - {\bf p}'_i )  \,.
\ee
Note that this normalisation is different than that of $\ket{\tildepin}$. 

The projection operator, ${\cal P}_x$, implementing our selective measurement, is a part of this complete set
\be
{\cal P}_x = \sum_{_{A,B}} 
\int_x d \Pi_{\gamma \delta} | p_f ; A,B \ketbra p_f ; A,B |\,,
\label{proj}
\ee
where $\ket{p_f ; A,B}$ is the $\gamma \delta$ final state with the definite momenta, collectively denoted by $p_f$, and spins/flavours, $x$ represents the selected momentum region
and $d \Pi_{\gamma \delta} = d \Pi_{\gamma} d \Pi_{\delta}$. 
With this event selection,
the evolved state $\roSE$ is projected to 
\bea
\roSE ~\mapsto~ \varrho'_x &=&
\Tr_P \left[ {\cal P}_x \roSE {\cal P}_x \right]
\nonumber \\
&=&
\sum_{_{I,J,K,L}} \sum_{_{A,B,C,D}} {\rin}_{[I,J],[K,L]}
\int_x d \Pi_{\gamma \delta}  \,
 \bra{p_f ; A,B} 
 S \ket{ \tildepin ; I,J } 
\times
 \nonumber \\
 &&~~~~~~~~~~~~~
  \times \bra{ \tildepin ; K,L } S^\dagger 
 \ket{p_f ; C,D}  | A,B \ketbra C,D | 
 \,,
\label{QI}
\eea
where $\Tr_P$ is the partial trace over the final state momenta.
We note that the projector \eqref{proj} involves the selection of $\gamma \delta$ out of all possible final states, so it implements a selective measurement even if $x$ is the full range of momenta.

\end{description}

The above set of steps directly corresponds to the procedure of the quantum instrument described in 
Section \ref{sec:review}.
The map $\rin \mapsto \varrho'_x$ is linear and trace non-increasing.  
However, as we demonstrate in the Appendix \ref{app}, $\varrho_x'$ is proportional to a dimensionful factor, which vanishes in the limit of infinite time and volume.  
To improve the situation, we rescale $\varrho_x'$ by some factor, which is formally singular and independent of $\rin$, to define 
\bea
\varrho_x &=&
\frac{1}{\sigma_{\cal N}}
\sum_{_{I,J,K,L}} \sum_{_{A,B,C,D}} {\rin}_{[I,J],[K,L]}
{R_x}^{[I,J],[K,L]}_{[A,B],[C,D]}
\, | A,B \ketbra C,D |\,.
\label{varrho_form}
\eea
Though the detailed derivation of this expression is provided in Appendix \ref{app}, the factors $R_x$ and $\sigma_{\cal N}$ need to be defined. 
First, $R_x$ is given (neglecting the mass of colliding particles) by
\be
{R_x}^{[I,J],[K,L]}_{[A,B],[C,D]}
\,\equiv\,
\frac{1}{2 s}
\int_x d \Pi_{\rm LIPS}   
 {\cal M}^{I,J}_{A,B} 
 \left( {\cal M}^{K,L}_{C,D} \right)^* \,,
\label{Rdef}
\ee
in the centre of mass frame, where $s$ is the square of the centre of mass energy, ${\cal M}^{I,J}_{A,B} \propto \bra{p_f ; A,B} S \ket{\tildepin ; I,J}$ is the scattering amplitude 
and 
\be\label{LIPS}
d \Pi_{\rm LIPS} \,=\, 
(2 \pi)^4 \delta^4 \left( \sum \pin^\mu - \sum p_{f}^\mu  \right)
\prod_{j=\gamma,\delta} \frac{d^3 p_j}{(2 \pi)^3} \frac{1}{2 E_j}
\ee
is the Lorentz invariant phase space measure of the final state. The quantity 
${R_x}^{[I,J],[K,L]}_{[A,B],[C,D]}$ is related to the cross section 
\be
\sum_{_{A,B}} {R_x}^{[I,J],[I,J]}_{[A,B],[A,B]} = \sigma_x(\alpha \beta[I,J] \to \gamma \delta) \,,
\ee
where the initial internal state is $\ket{I,J} \in \Hin$, and the final state momenta are restricted in $x$, but the final state spins/flavours are unconstrained. 

For convenience, we choose the normalisation factor $\sigma_{\cal N}$ to be
the inclusive cross section of $\alpha \beta \to \gamma \delta$ 
for the maximally mixed initial state, $\rin^{\rm mix} \equiv \frac{1}{\din} \sum_{IJ} |I,J \ketbra I,J|$,
\be
\sigma_{\cal N} \,=\, \sigma(\alpha \beta [\rin^{\rm mix}] \to \gamma \delta),
\label{sigN}
\ee
where the final state momenta and spins/flavours are unconstrained.
The rescaling from $\varrho_x'$ to $\varrho_x$ is independent of the initial state $\rin$, therefore the map $\rin \mapsto \varrho_x$ is still linear. Because the normalisation factor was introduced by hand, the quantity
\be
\Tr \varrho_x \,=\,
\frac{\sigma_x(\alpha \beta[\rin] \to \gamma \delta)}{\sigma(\alpha \beta[\rin^{\rm mix}] \to \gamma \delta )} \,,
\label{tr_sim}
\ee
 is not necessarily bounded by 1. 
%
Nevertheless, the final spin state $\rho_x \in {\cal S}({\cal H}_{\gamma \delta})$ is given by formula \eqref{poststate}, as in the standard quantum instrument, so that the artificial normalisation factor cancels out.

In summary, the outcome of the quantum instrument at a collider measurement, ${\cal I}_x(\rin)$, is given by Eqs. (\ref{varrho_form}--\ref{LIPS}) and (\ref{sigN}).
Using the linearity of the map and Eq. \eqref{expand},
we find that the rescaled Choi matrix element is given by
\be
{\cal I}_x ( |I,J \ketbra K,L | )_{[A,B],[C,D]} ~=~
\frac{1}{d_{\rm in}}
\frac{1}{\sigma_{\cal N}} \frac{1}{2 s}
\int_x d \Pi_{\rm LIPS} \,
 {\cal M}^{I,J}_{A,B} \left( {\cal M}^{K,L}_{C,D} \right)^* \,.
 \label{IxTheo_0}
\ee

Finally, we note that $\varrho_x$ can be obtained from $\rin$ with the help of rescaled Kraus operators as
\be\label{IKrauss}
\varrho_x =  \mathcal{I}_x(\rin) = \int_x d \Pi_{\rm LIPS} E(p_f) \rin E^\dagger(p_f),
\ee
with
\be
E(p_f) = \frac{1}{\sqrt{2 s \sigma_{\cal N}}}
\sum_{_{I,J,A,B}} {\cal M}_{A,B}^{I,J} | A,B \ketbra I,J | \,.
\ee

For any given initial state $\rin = \sum_{_{I,J,K,L}} {\rin}_{[I,J],[K,L]}  | I,J \ketbra K,L|$, the Choi matrix \eqref{IxTheo_0}, calculated within a specific model, enables us to immediately calculate the model's prediction of the outcome as 
\be
\bra{A,B} \varrho_x \ket{C,D} \,=\,  
d_{\rm in} \sum_{_{I,J,K,L}} ({\rin})_{[I,J],[K,L]} \, {\cal I}_x \left( | I,J \ketbra K,L| \right)_{[A,B],[C,D]} \,.
\label{Ix_pred}
\ee

\subsection{Foundational tests via Choi matrix reconstruction}

An experimental measurement of the elements of the Choi matrix for a scattering process offers a unique possibility to test the predictions of a particular quantum field theoretic model (e.g. Standard Model), and of quantum theory in general.

The possibility of performing the relevant Choi matrix reconstruction is based on three assumptions:

\begin{enumerate}

    \item The initial spins and/or flavours of both particles $\alpha$ and $\beta$ can be prepared independently in any of the $\din^2$ linearly independent states of the form 
 $\rin^{(a,b)} = \rho_{\ini,\alpha}^a \otimes \rho_{\ini,\beta}^b$.


    \item For any input state $\rin^{(a,b)}$ one can perform a reliable quantum state tomography which yields a reconstructed state $\rho_x^{(a,b)} \in S(\Hout)$ of the spins and/or flavours of the outgoing particles $\gamma$, $\delta$, for some range $x$ of their kinematic parameters.

    \item One can measure:

    \begin{itemize}
        \item $\sigma\big(\alpha \beta [\rin^{\rm mix}] \to \gamma \delta\big)$ --- the inclusive cross section for the process $\alpha\beta \to \gamma \delta$ with an ensemble of random spins and/or flavours of $\alpha$ and $\beta$;
        \item $\sigma_x\big( \alpha \beta [\rin^{(a,b)}] \to \gamma \delta \big)$ --- the effective  
cross section after the kinematic selection, $x$, of the $\gamma \delta$ momenta, for any of the initial states $\rin^k$.
    \end{itemize}
\end{enumerate}

With the above assumptions, one obtains $d_{\rm in}^2$ quantum instrument outcomes:
\be
\varrho^{(a,b)}_x \,=\, 
\frac{\sigma_x\big( \alpha \beta [\rin^{(a,b)}] \to \gamma \delta \big)}{\sigma\big(\alpha \beta [\rin^{\rm mix}] \to \gamma \delta\big)}
\cdot \rho^{(a,b)}_x\,.
\label{rho2varpho}
\ee
The first condition implies that $\rho_{{\rm in},\alpha}^a$ and $\rho_{{\rm in},\beta}^b$ span the spaces ${\cal S(H_\alpha)}$ and ${\cal S(H_\beta)}$, respectively. Hence, we can find the coefficients $A_a^{(I,K)}$ and $B_b^{(J,L)}$ satisfying 
\be
| I \ketbra K |_{\alpha} \,=\, \sum_{a=1}^{{\rm dim} {\cal H}_\alpha} 
A_{a}^{(I,K)} \, \rho_{{\rm in},\alpha}^a
~~~{\rm and}~~~
| J \ketbra L |_{\beta} \,=\, \sum_{b=1}^{{\rm dim} {\cal H}_\beta} 
B_{b}^{(J,L)} \, \rho_{{\rm in},\beta}^b\,.
\label{AB}
\ee
Using the linearity of the quantum instrument map, the Choi matrix can be reconstructed from the experimental data --- through the quantum process tomography --- as follows (cf.\ Eqs.\ \eqref{TheNewX} and \eqref{TheNewX2})
\be
{\cal I}_x ( |I,J \ketbra K,L | )_{[A,B],[C,D]}
\,=\, 
\frac{1}{d_{\rm in}} 
\sum_{a=1}^{{\rm dim} {\cal H}_\alpha}
\sum_{b=1}^{{\rm dim} {\cal H}_\beta} 
A_{a}^{(I,K)} 
B_{b}^{(J,L)}  \cdot 
\bra{A,B} \varrho^{(a,b)}_x \ket{C,D} \,.
\label{Choi_exp}
\ee
A more formal discussion on the quantum process tomography is given in Appendix \ref{app:tomo}.

\begin{figure}[t!]
\begin{center}
\includegraphics[scale=1.5]{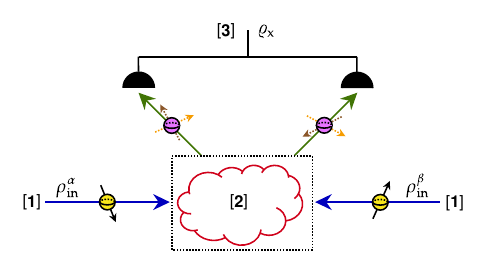}
\caption{\label{fig:Qscatt}Experimental reconstruction of the Choi matrix associated with 2 $\to$ 2 scattering process. At step [\textbf{1}] we prepare the initial state, $\rin^{\alpha} \otimes \rin^\beta$, of the two beams. Step [\textbf{2}] is the scattering process, which is treated as a quantum-information-processing `black-box'. At step [\textbf{3}]  we reconstruct the relevant (unnormalised) states $\varrho_x$ from the gathered data. The dotted coloured arrows of the outgoing particles illustrate the fact their joint spin (or flavour) state is entangled.}
\end{center}
\end{figure}

What distinguishes the Choi matrix reconstruction from typical high-energy tests is that the computation of formula \eqref{Choi_exp} from the data is model-independent and, in fact, it does not rely on the quantum description of the scattering processes. Indeed, formula \eqref{Choi_exp} only relies on steps {\bf [1]} and {\bf [3]} of the procedure described in Sec. \ref{sec:S}, but {\it not} on step {\bf [2]} (evolution). In other words, while we do assume the validity of quantum theory at the stages of state preparation and measurement, we treat the $\alpha\beta \to \gamma\delta$ scattering process as a `black box',  which need not abide by the laws of quantum theory. 
This point and the procedure of quantum process tomography at colliders are illustrated in Fig.\ \ref{fig:Qscatt}.
The experimental Choi matrix reconstruction is thus a ``Q-data test'' --- see \cite{Eckstein:2021pgm} for more details.

A Choi matrix reconstructed from experimental data carries a lot of useful information about the scattering process $\alpha \beta \to \gamma \beta$. It can be directly compared with the theoretical prediction \eqref{IxTheo_0} based on a specific model. Note that the dimension of the Choi matrix associated with a quantum instrument is typically fairly large --- even for a simple two-qubit to two-qubit mapping it has $(\din \cdot \dout)^2 = 256$
entries. A disagreement between the SM prediction and the experimental value at {\it any} of these entries would hint at new physics. Therefore, the experimental Choi matrix reconstruction is a powerful tool for a precision test of the Standard Model, as well as for testing specific BSM models.

The black-box approach to high-energy scattering processes offers an unprecedented opportunity to explore physics beyond the perturbative Standard Model. In particular, the Choi matrix \eqref{Choi_exp} could be measured for processes in hadron physics, which are not calculable through Feynman diagrams. This would provide a new insight into the non-perturbative dynamics of quarks and gluons, which could be compared with lattice QCD simulations. 

Even more importantly, the Choi matrix reconstruction is a foundational test of the quantum theory itself, akin to Bell tests. Indeed, as explained in Section \ref{sec:review}, quantum mechanics postulates that the map ${\cal I}_x: \rin \mapsto \varrho_x$ is linear and completely positive, for any range $x$ of kinematical parameters. By the channel-state duality, this is equivalent to the Choi matrix $\widetilde{\cal I}_x$ being positive semi-definite.  
This property of the map ${\cal I}_x$ can be checked directly by inspecting the smallest eigenvalue of the experimentally reconstructed Choi matrix \eqref{Choi_exp},
\begin{align}
    \lambda_\text{min}(\widetilde{\cal I}_x) = \min \big\{ \lambda \, \big\vert \, \det \big(\widetilde{\cal I}_x - \lambda \big) = 0 \big\}.
\end{align}
If the experimental data suggest a statistically significant deviation from positivity, $\lambda_\text{min}(\widetilde{\cal I}_x) \geq  0$, then this would imply that the scattering process cannot be entirely explained by {\it any} quantum model.

Furthermore, the postulated linearity of quantum theory implies that the action of the map ${\cal I}_x: \rin \mapsto \varrho_x$ on basis states completely determines its outcome for any other state, 
Consequently, one can directly test the linearity postulate as follows:

Suppose that we have gathered a dataset $\{\rin^k,\varrho_x^k\}_{k=1}^K$, with $K > \din^2$, which consists of pairs of the prepared input states $\rin^k$ and the corresponding operators $\varrho_x^k$ reconstructed from quantum state tomography and cross-sections. Out of this dataset we choose $\din^2$ entries with linearly independent $\rin$'s (we can arrange them to be the first $\din^2$ entries) and write down from formula \eqref{Choi_exp} the block-matrices ${\cal I}_x \left( | I,J \ketbra K,L| \right)$, which in turn determine the output operators
\begin{align}\label{lin_test}
   \xi_x^k \equiv {\cal I}_x (\rin^k) = \sum_{_{I,J,K,L}} ({\rin^k})_{[I,J],[K,L]} \, {\cal I}_x \left( | I,J \ketbra K,L| \right),
\end{align}
for all $k > \din^2.$ If the scattering process induces a linear map from the space ${\cal S}(\Hin)$ to ${\cal B}_+(\Hout)$, then we should have $\xi_x^k = \varrho_x^k$ for all $k$. This is a \emph{prediction} of quantum theory, which can be verified empirically.
The test can be performed for each Choi matrix element and any input state $\rin^k$ with $k > d_{\rm in}^2$ by comparing the deviation between the prediction and the outcome to their uncertainties:
\be
\left| \frac{  \bra{A,B} \varrho_x^k - \xi_x^k \ket{C,D}  }{    \sigma^k_{[A,B],[C,D]}  } \right|^2\,,
\ee
where $\sigma^k_{[A,B],[C,D]}$ is the combined uncertainty including the theoretical, statistical and systematic uncertainties for the measurement and calculation of the $[A,B],[C,D]$ component of $\varrho_x^k$ and $\xi_x^k$.
If one finds a statistically significant deviation between the prediction and outcome in any of these tests, it would be a strong indication that the scattering process involves some nonlinear dynamics of quantum information, which cannot be accommodated within the standard quantum theory.

\section{An example: polarised $e^- e^+ \to t \bar t$ scattering}
\label{sec:tomography}

Let us now illustrate the procedures on a concrete process involving a map between the spin density matrices of two ingoing and outgoing fermions. 
To this end, we need a process in which one can control the polarisation of the colliding particles, reconstruct the spins of the final-state particles, and in which the final state depends non-trivially on the initial state polarization. 
As an illustrative example, we consider the top-quark pair production at a lepton collider 
\be
e^- e^+ \to t \bar t
\ee
and study the transition of the spin states. 
The initial spin state of the electron-positron composite system is a product state $\rin = \rin^{e^-} \otimes \rin^{e^+} \in {\cal S}(\H_\text{in})$, where $\H_\text{in} = \C^2_{e^-} \otimes \C^2_{e^+}$, as the two beams are not initially correlated and are polarised independently. On the other hand,
$\rho_x^{t\bar{t}} \in {\cal S}(\H_\text{out})$, with $\H_\text{out} = \C^2_{t} \otimes \C^2_{\bar t}$, represents the effective final spin state of the top-antitop system for a given range $x$ of the $t\bar{t}$ momentum.

We work in the centre of mass frame and define the $z$-axis along the electron beam.  For concreteness, the $y$-axis is defined as being in the opposite direction of the Earth's centre. 
We quantise the electron and positron spins in the $z$ direction.
The simultaneous eigenstates of the $e^- e^+$ spins are denoted by
$\{ \ket{I,J} \} = \{ \ket{++}, \ket{+-}, \ket{-+}, \ket{--} \}$
where in the last
expression, the first and second signs represent the eigenvalues of $\hat S_z^{e^-}$ and $\hat S_z^{e^+}$, respectively. 
To describe the $t \bar t$ spins, it is convenient to work in the helicity basis, in which the three unit vectors (axes) $\{ {\bf r}, {\bf n}, {\bf k} \}$ are defined as follows: ${\bf k}$ is taken to be the direction of the top quark.
${\bf r} \equiv ({\bf z} - {\bf k} \cos \theta)/\sin \theta$ lies in the plane spanned by ${\bf k}$ and the $z$-axis and ${\bf r}$ points the beam direction, where $\theta$ is the angle between ${\bf k}$ and $\bf z$.
Finally, ${\bf n}$ is defined as ${\bf n} \equiv {\bf k} \times {\bf r}$.
The top and antitop spins are quantised in the ${\bf k}$ direction.
The simultaneous eigenstates of the $t$ and $\bar t$ spins are denoted by
$\{ \ket{A,B} \} = \{ \ket{00}, \ket{01}, \ket{10}, \ket{11} \}$.

\subsection{Theoretical calculation}
\label{sec:prediction}

In this section, we derive a theoretical prediction for the Choi matrix for polarised $e^-e^+ \to t\bar{t}$ scattering through a tree-level calculation in quantum field theory.
For generality, we use the following effective interaction Lagrangian:
\be
{\cal L} ~\ni~ 
\sum_i \frac{1}{\Lambda^2_i}
[ \bar \psi_e \gamma_\mu (c_L^i P_L + c_R^i P_R) \psi_e ]
[ \bar \psi_t \gamma^\mu (d_L^i P_L + d_R^i P_R) \psi_t ]\,,
\label{L}
\ee
where $i$ runs over possible contributions.
For example, the Standard Model contribution can be obtained by summing over the photon and $Z$-boson exchange, $i = A$ and $Z$, with the following replacement at the amplitude level:
\vskip2mm
\begin{center}
\renewcommand{\arraystretch}{1.3}
\begin{tabular}{ c | c c c c c}
$i$ & $\Lambda^2_i$ & $c_L^i$ & $c_R^i$ & $d_L^i$ & $d_R^i$ \\ 
\hline
$A$ & $s$ & $-e$ & $-e$ & $\tfrac{2}{3} e$ & $\tfrac{2}{3} e$  \\ 
$Z$ & $s-m_Z^2 + i m_Z \Gamma_Z$ 
& $g_Z \left( -\tfrac{1}{2} + \sin^2 \theta_w \right)$ 
& $g_Z \sin^2 \theta_w$
& $g_Z \left( \tfrac{1}{2} - \tfrac{2}{3} \sin^2 \theta_w \right)$
& $g_Z \left( - \tfrac{2}{3} \sin^2 \theta_w \right)$
\end{tabular}
\end{center}
\vskip2mm
Here $\theta_w$ is the weak mixing angle, 
$m_Z$ and $\Gamma_Z$ are the mass and width of the $Z$ boson, and $e$ and $g_Z = e/(\sin \theta_w \cos \theta_w)$ are the electromagnetic and the $Z$-boson couplings, respectively.

We perform the calculation in the centre of mass frame and parametrise the top-quark momentum, $p^{\mu}_t = (E, q \sin \theta \cos \phi, q \sin \theta \sin \phi, q \cos \theta)$ with $0 \le \theta \le \pi$ and $- \pi \le \phi \le \pi$, and
$E = \sqrt{m_t^2 + q^2} = \tfrac{\sqrt{s}}{2}$. The two kinematic parameters $\theta$ and $\phi$ determine our range $x$ of classical outcomes of the momentum measurement, $x \subset [0, \pi] \times  [-\pi, \pi]$.
In this setup, the tree-level helicity amplitudes are given by

\begin{eqnarray}
 {\cal M}^{++}_{00} ~=~ {\cal M}^{++}_{11} &=& 
e^{i \phi} \sum_i \frac{s}{2 \Lambda^2_i}
\gamma^{-1} c^i_R \sin \theta ( d^i_L + d^i_R )\,,
\nonumber \\
 {\cal M}^{++}_{01} &=&
- e^{i \phi} \sum_i \frac{s}{2 \Lambda^2_i} c^i_R (1 + \cos \theta) 
 [ 
d^i_L (1-\beta) + d^i_R (1+\beta) ]\,,
\nonumber \\
{\cal M}^{++}_{10} &=& 
- e^{i \phi} \sum_i \frac{s}{2 \Lambda^2_i} c^i_R (1 - \cos \theta)
 [ 
d^i_L (1 + \beta) + d^i_R (1 - \beta) 
]\,,
\nonumber \\
{\cal M}^{--}_{00} ~=~ {\cal M}^{--}_{11} &=&
e^{-i \phi} \sum_i \frac{s}{2 \Lambda^2_i} \gamma^{-1} c^i_L \sin \theta 
 (d^i_L + d^i_R)\,,
\nonumber \\
{\cal M}^{--}_{01} &=&
- e^{-i \phi} \sum_i \frac{s}{2 \Lambda^2_i} c^i_L (1 - \cos \theta)  [ 
d^i_L (1-\beta) 
+ d^i_R (1+\beta) 
]\,,
\nonumber 
\\
{\cal M}^{--}_{10} &=&
- e^{-i \phi} \sum_i \frac{s}{2 \Lambda^2_i} c^i_L (1 + \cos \theta) 
 [ 
d^i_L (1+\beta) 
+ d^i_R (1-\beta) 
]\,,
\label{Ms}
\end{eqnarray}

where $\beta = \frac{q}{E}$ and $\gamma = \frac{E}{m_t}$.
For other $e^- e^+$ polarisations, the amplitudes vanish:
${\cal M}^{+-}_{[A,B]} = {\cal M}^{-+}_{[A,B]} = 0$.
The phase space factor is
\be
d \Pi_{\rm LIPS} ~=~ \frac{1}{16 \pi^2}\frac{q}{ \sqrt{s} } d \Omega \,,
\ee
with $d \Omega = \sin \theta d \theta d \phi$. 
The Choi matrix elements $\choiI_x = \int_x d \choiI$
are obtained through (see Eq.\ \eqref{IxTheo_0})
\bea
d {\cal I} ( |I,J \ketbra K,L | )_{[A,B],[C,D]} 
\,=\,
\frac{1}{128 \pi^2} 
\frac{1}{\sigma_{\cal N}}  \frac{q}{ s \sqrt{s}} \, {\cal M}^{I,J}_{A,B} \left( {\cal M}^{K,L}_{C,D} \right)^*\!
d \Omega\,,
\label{IxTheo}
\eea
and the normalisation factor \eqref{sigN} is given by
\bea
\sigma_{\cal N} &=& \frac{1}{128 \pi^2} \frac{q}{s \sqrt{s}} \int  \sum_{I,J,A,B} \left| {\cal M}^{I,J}_{A,B} \right|^2
d \Omega
\,.
\label{sigNee}
\eea

The spin structure of the interaction Lagrangian \eqref{L} implies that, at tree-level within the Standard Model, the Choi matrix $\choiI_x$ has the following form
\begin{align}\label{Choi_eett}
\choiI_x \,=\,
\frac{1}{4} 
\begin{pmatrix}
{\cal I}_x(|++ \ketbra ++|)  & 0 & 0 & {\cal I}_x(|++ \ketbra --|) \\ 
0 & 0 & 0 & 0 \\ 
0 & 0 & 0 & 0 \\ 
{\cal I}_x(|-- \ketbra ++|)  & 0 & 0 & {\cal I}_x(|-- \ketbra --|)  
\end{pmatrix},
\end{align}
where each ${\cal I}_x(|IJ \ketbra KL|) $ is a $4 \times 4$ matrix. By Hermicity of $\choiI_x$ we have
${\cal I}_x(|-- \ketbra ++|)  = [ {\cal I}_x(|++ \ketbra --|)  ]^\dagger$.

The angular dependence, $d \choiI / d \Omega$ of the Choi matrix for polarised $e^-e^+ \to t \bar{t}$ scattering can be calculated explicitly using formulas  Eqs.\ \eqref{Ms}, \eqref{IxTheo} and \eqref{sigNee}. With $s_\theta \equiv \sin \theta$ and $c_\theta \equiv \cos \theta$, we obtain
\bea
\frac{d {\cal I}(|++ \ketbra ++|) }{d \Omega} 
&=&
\begin{pmatrix}
a^{(+)}_{11} s^2_\theta & a^{(+)}_{12} s_\theta (1 + c_\theta) & a^{(+)}_{13} s_\theta (1 - c_\theta) & a^{(+)}_{14} s^2_\theta \\
a^{(+)}_{21} s_\theta (1 + c_\theta) & a^{(+)}_{22} (1 + c_\theta)^2 & a^{(+)}_{23} s^2_\theta & a^{(+)}_{24} s_\theta (1 + c_\theta) \\
a^{(+)}_{31} s_\theta (1 - c_\theta) & a^{(+)}_{32} s^2_\theta & a^{(+)}_{33} (1 - c_\theta)^2 & a^{(+)}_{34} s_\theta (1 - c_\theta) \\
a^{(+)}_{41} s^2_\theta & a^{(+)}_{42} s_\theta (1 + c_\theta) & a^{(+)}_{43} s_\theta (1 - c_\theta) & a^{(+)}_{44} s^2_\theta 
\end{pmatrix} \,,
\nonumber \\
\frac{d {\cal I}(|++ \ketbra --|) }{d \Omega} 
&=&
e^{i 2 \phi}
\begin{pmatrix}
a^{(+-)}_{11} s^2_\theta & a^{(+-)}_{12} s_\theta (1 - c_\theta) & a^{(+-)}_{13} s_\theta (1 + c_\theta) & a^{(+-)}_{14} s^2_\theta \\
a^{(+-)}_{21} s_\theta (1 + c_\theta) & a^{(+-)}_{22} s^2_\theta & a^{(+-)}_{23} (1 + c_\theta)^2 & a^{(+-)}_{24} s_\theta (1 + c_\theta) \\
a^{(+-)}_{31} s_\theta (1 - c_\theta) & a^{(+-)}_{32} (1 - c_\theta)^2 & a^{(+-)}_{33} s^2_\theta & a^{(+-)}_{34} s_\theta (1 - c_\theta) \\
a^{(+-)}_{41} s^2_\theta & a^{(+-)}_{42} s_\theta (1 - c_\theta) & a^{(+-)}_{43} s_\theta (1 + c_\theta) & a^{(+-)}_{44} s^2_\theta 
\end{pmatrix}\,,
\nonumber \\
\frac{d {\cal I}(|-- \ketbra --|) }{d \Omega} 
&=&
\begin{pmatrix}
a^{(-)}_{11} s^2_\theta & a^{(-)}_{12} s_\theta (1 - c_\theta) & a^{(-)}_{13} s_\theta (1 + c_\theta) & a^{(-)}_{14} s^2_\theta \\
a^{(-)}_{21} s_\theta (1 - c_\theta) & a^{(-)}_{22} (1 - c_\theta)^2 & a^{(-)}_{23} s^2_\theta & a^{(-)}_{24} s_\theta (1 - c_\theta) \\
a^{(-)}_{31} s_\theta (1 + c_\theta) & a^{(-)}_{32} s^2_\theta & a^{(-)}_{33} (1 + c_\theta)^2 & a^{(-)}_{34} s_\theta (1 + c_\theta) \\
a^{(-)}_{41} s^2_\theta & a^{(-)}_{42} s_\theta (1 - c_\theta) & a^{(-)}_{43} s_\theta (1 + c_\theta) & a^{(-)}_{44} s^2_\theta 
\end{pmatrix}\,.
\label{Imatrix}
\eea
Here, the coefficients $a^{(\star)}_{ij}$ depend on the centre of mass energy and of the explicit form of the interaction Lagrangian \eqref{L}.   

The differential Choi matrix, $d \choiI / d \Omega$, involves a single Kraus operator, as can be seen in Eq.\ \eqref{IKrauss}. 
It, therefore, has only one non-zero eigenvalue as evidenced by Eqs.\ \eqref{choi_mix} and \eqref{EkO}.
This eigenvalue is equal to the trace of the matrix:
\be
\Tr \left[ d \choiI / d \Omega \right]  
~=~ \tfrac{1}{4} \left( D_0 + 2 D_1 \cos \theta + D_2 \cos^2 \theta \right),
\label{traceChoi}
\ee
with 
\bea
&& D_0 ~=~ \sum_{i=1}^4 \left[ a^{(+)}_{ii} + a^{(-)}_{ii} \right],
\nonumber \\
&& D_1 = a^{(+)}_{22} - a^{(+)}_{33} - a^{(-)}_{22} + a^{(-)}_{33},
\nonumber \\
&& D_2 = - a^{(+)}_{11} + a^{(+)}_{22} - a^{(+)}_{33} + a^{(+)}_{44} 
- a^{(-)}_{11} + a^{(-)}_{22} - a^{(-)}_{33} + a^{(-)}_{44}.
\eea

Observe that the integral of the off-diagonal block over a half-full (or full) range of the azimuthal angle $\phi$ vanishes. This means that in order to register a non-zero off-diagonal block of the Choi matrix \eqref{Choi_eett} we need to perform the reconstruction using events with finer range of the top quark azimuthal scattering angles.

In Appendix \ref{numerical}, we present the values of the coefficient $a^{(\star)}_{ij}$ evaluated at tree-level in the Standard Model for the reference centre of mass energies $\sqrt{s} = 370$ and 1000 GeV. 

\begin{figure}[h]
\begin{center}
\textbf{a)}\includegraphics[scale=0.5]{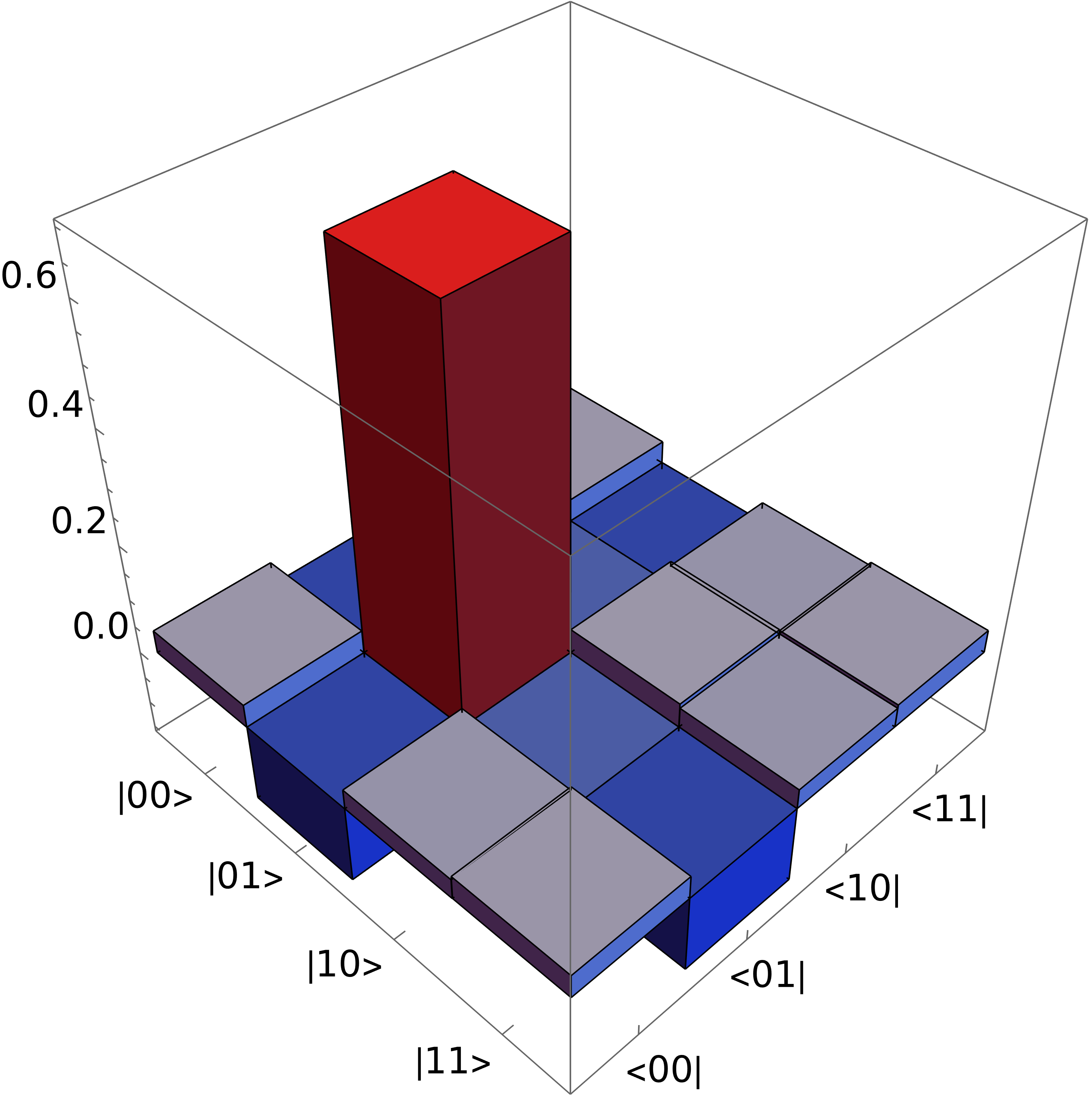}\; \textbf{b)}\includegraphics[scale=0.5]{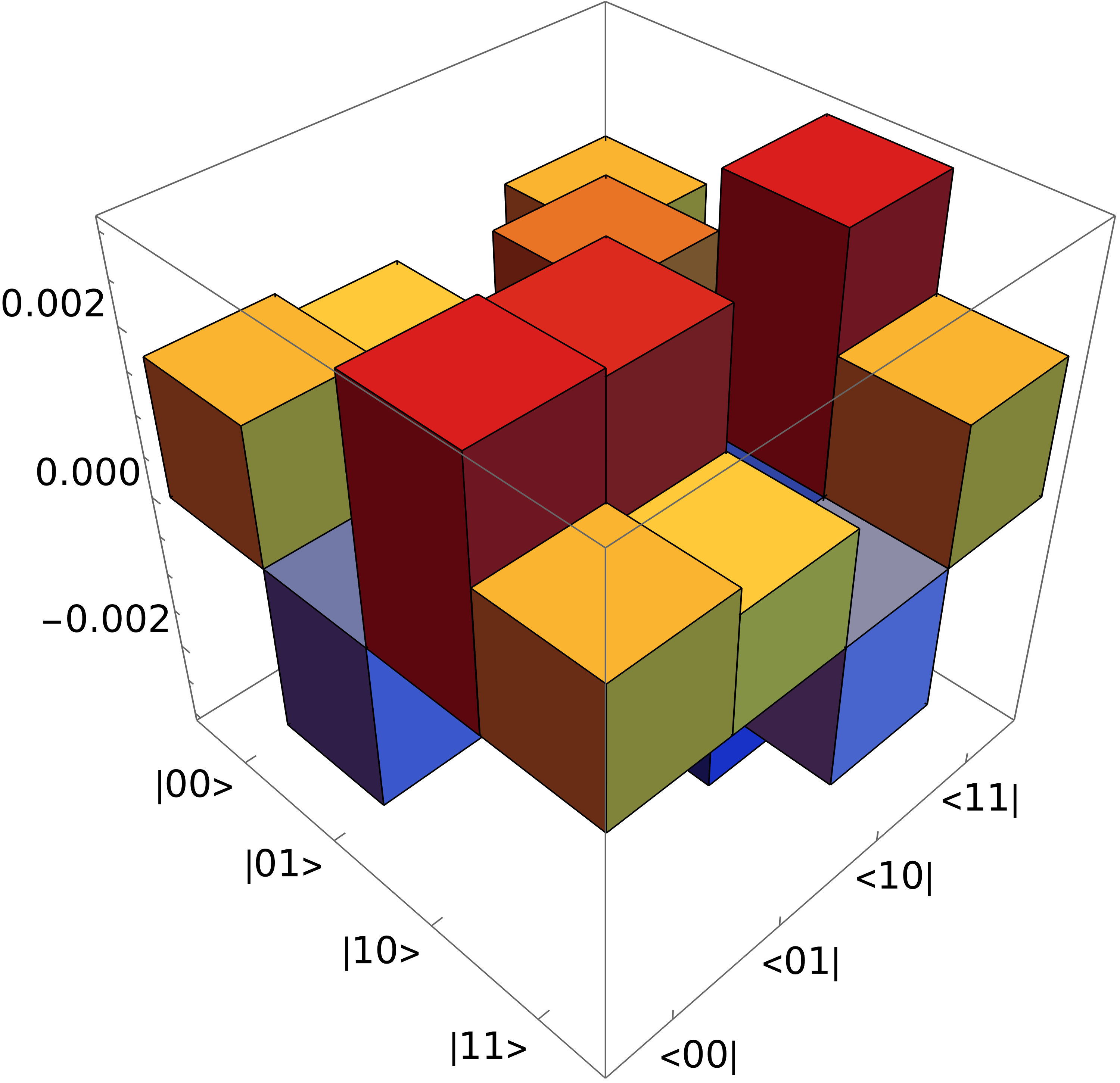}
\caption{\label{fig:tomo}An illustration of the block elements of the Choi matrix \eqref{Choi_eett} for $e^-e^+ \to t \bar{t}$ process evaluated at tree level in the Standard Model at center of mass energy $\sqrt{s} = 370$~GeV. \textbf{a)} The diagonal block ${\cal I}_x(|++ \ketbra ++|)$ for the full range of kinematic parameters, $x = \{ \theta \subset[0,\pi], \phi \subset [-\pi,\pi]\}$.  \textbf{b)} The off-diagonal block ${\cal I}_x(|++ \ketbra --|)$ for a restricted range of kinematic parameters, $x = \{ \theta \subset [2\pi/3,\pi] , \phi \subset [-\pi/4,\pi/4] \}$. 
}
\end{center}
\end{figure}

We illustrate two blocks of the Choi matrix evaluated at $\sqrt{s} = 370$~GeV in Fig.\ \ref{fig:tomo}.
On the left plot \textbf{a)}, the ${\cal I}_x(|++ \ketbra ++|)$ block with the unrestricted scattering angle, $x = \{ \theta \subset[0,\pi], \phi \subset [-\pi,\pi]\}$, is shown.
We see that the $|01 \ketbra 01|$ component clearly dominates, though some admixture of other elements, e.g. $| 01 \ketbra 00|$ or $| 11 \ketbra 01|$, is also visible.
The right plot \textbf{b)} illustrates the off-diagonal block ${\cal I}_x(|++ \ketbra --|)$ with a restricted scattering angle, $x = \{ \theta \subset [2\pi/3,\pi] , \phi \subset [-\pi/4,\pi/4] \}$.
Although the off-diagonal blocks in Eq.\ \eqref{Choi_eett} in general have complex entries,
all matrix entries are real with the above kinematic range $x$.
We see that the off-diagonal block of the Choi matrix \eqref{Choi_eett} has a very rich structure with all 16 elements substantially different from zero. However, their magnitudes are significantly smaller than the dominant component of ${\cal I}_x(|++ \ketbra ++|)$.

\subsection{Experimental procedure}\label{sec:ee_exp}

Our goal is to reconstruct the $16 \times 16$ Choi matrix \eqref{Choi_exp} from the experimental data. To this end we need to perform 16 experimental runs with 16 distinct pairs of 4 different polarisations of the electron beam and 4 different polarisations of the positron beam. Polarised fermion beams have been used in colliders since the 1970s \cite{Buon:1993rtf}. While most of the experiments featured fermion beams polarised along the beam axis, transverse polarisation of the electron/positron beam is also possible \cite{sobloher2012polarisationherareanalysis}.  The practicalities and capabilities of current and future colliders facilitating beam polarization are discussed further in Section~\ref{sec:practicalities}.

A general polarisation vector of a spin-half fermion can be written as
\bea
\ket{\bf n} ~=~ \alpha \ket{+} + \beta \ket{-},
\label{pure_pol}
\eea
with $|\alpha|^2 + |\beta|^2 = 1$ and $\ket{\pm}$ denoting the eigenstates of the spin operator along a chosen axis. Our choice for the $e^- e^+ \to t \bar t$ process is the $z$-axis, along the electron beam. We could write the coefficients in \eqref{pure_pol} as  $\alpha = \cos (\theta^\prime/2)$, $\beta = e^{i \phi^\prime} \sin (\theta^\prime/2)$, which shows that a perfectly polarised spin-half fermion beam corresponds to a pure qubit state on the Bloch sphere. In practice, the polarisation is never perfect and we have to model the initial spin state of a fermion in the beam in terms of a mixed state inside the Bloch ball. Concretely, if a beam of spin-half fermions is polarised along axis ${\bf n}$ at degree $100 \cdot q$\,\%, then the corresponding spin density matrix is given by 
\bea
\label{beam_pol}
\eta(q,{\bf n}) &\equiv& q\, |{\bf n} \ketbra {\bf n} | + \tfrac{1 - q}{2}\, {\bf 1}
\nonumber \\
&=&
\tfrac{1}{2} \big[ (1 + q) | {\bf n} \ketbra {\bf n} | + (1-q) |{-\bf n} \ketbra -{\bf n} | \big]
\nonumber \\
&=& \sum_{I,K = \pm} C^{(I,K)}_{(q,{\bf n})} \,| I \ketbra K |\,
\label{eta_in}
\eea
with
\be
\left(
C^{(+,+)}_{(q,{\bf n})}, ~
C^{(+,-)}_{(q,{\bf n})}, ~
C^{(-,+)}_{(q,{\bf n})}, ~
C^{(-,-)}_{(q,{\bf n})} 
\right)
=
\big(
q |\alpha|^2 + \tfrac{1-q}{2}, ~
q \alpha \beta^*, ~
q \alpha^* \beta, ~
q |\beta|^2 + \tfrac{1-q}{2} 
\big)\,.
\ee
In the last line of Eq.\ \eqref{eta_in}, we have used Eq.\ \eqref{pure_pol} and ${\bf 1} = |+ \ketbra +| + |- \ketbra -|$.


Two of the initial polarisations can be chosen to be along the beam, for the other two let us conveniently take the $\vec{x}$ direction and the $\vec{y}$ direction. We assume, for sake of simplicity, that the degree of polarisation of the beam does not depend on the direction (for the general case see Appendix \ref{app:tomo}). On the other hand, the degree of polarisation of the positron beam is typically lower than the one of the electron beam and we denote the former as $\bar{q}$. In summary, we prepare 16 different initial states
\begin{align}
\rin^{(a,b)} = \rho_{\text{in},e^-}^{a} \otimes \rho_{\text{in},e^+}^{b} \,,
\end{align}
with
\begin{align}
\rho_{\text{in},e^-}^{1} = \eta(q,+), \quad \rho_{\text{in},e^-}^{2} = \eta(q,-), \quad \rho_{\text{in},e^-}^{3} = \eta(q,\vec{x}), \quad \rho_{\text{in},e^-}^{4} = \eta(q,\vec{y})\,,
\nonumber \\
\rho_{\text{in},e^+}^{1} = \eta(\bar q,+), \quad \rho_{\text{in},e^+}^{2} = \eta(\bar q,-), \quad \rho_{\text{in},e^+}^{3} = \eta(\bar q,\vec{x}), \quad \rho_{\text{in},e^+}^{4} = \eta(\bar q,\vec{y})\,
\label{ini_e}
\end{align}
and $\ket{ \vec{x} } = \tfrac{1}{\sqrt{2}} \big(\ket{+} + \ket{-}\big)$ and 
$\ket{ \vec{y} } = \tfrac{1}{\sqrt{2}} \big(\ket{+} + i \ket{-}\big)$.

In this case, the input states are expanded in terms of the basis operators as
\be
\rho_{{\rm in},e^-}^a = \sum_{I,K = \pm} [A^{-1}]^a_{(I,K)} | I \ketbra K |,
~~~{\rm and}~~~
\rho_{{\rm in},e^+}^b = \sum_{J,L = \pm} [B^{-1}]^b_{(J,L)} | J \ketbra L |,
\ee
with
\be
A^{-1}
\,=\,
\frac{1}{2}
\begin{pmatrix}
1+q & 0 & 0 & 1-q \\
1-q & 0 & 0 & 1+q \\
1 & q & q & 1 \\
1 & -iq & iq & 1 
\end{pmatrix},
\ee
where the rows are indexed by $a \in \{1,2,3,4\}$ and the columns by pairs $(I,K) \in \{(++),(+-),(-+),(--)\}$.
Taking its inverse, we find the matrix appearing in Eq.\ \eqref{AB},
\be
A
\,=\,
\frac{1}{2q}
\begin{pmatrix}
1+q & -(1-q) & 0 & 0 \\
-(1+i) & -(1+i) & 2 & 2i \\
-1+i & -1+i & 2 & -2i \\
-(1-q) & 1+q & 0 & 0 
\end{pmatrix},
\ee
where now the columns are indexed by $a$ and the row by pairs $(I,K)$.
The matrix $B$ 
for the positron states has the same form with $q$ replaced by $\bar q$.

For each of the 16 initial states we reconstruct the corresponding output operators $\varrho_x^{(a,b)} = {\cal I}_x\big(\rin^{(a,b)} \big)$, with the chosen range $x$ of the $t \bar{t}$ kinematic parameters. This is done from the measurements via formula \eqref{rho2varpho},
\begin{align}
\varrho_x^{(a,b)} = \frac{\sigma_x\big( e^- e^+ [\rin^{(a,b)}] \to t \bar t \big) }{\sigma\big(\alpha \beta [\rin^{\rm mix}] \to t \bar t \big)} \, \rho_x^{(a,b)},
\end{align}
where $\sigma(e^- e^+ [\rin^{\rm mix}] \to t \bar t )$ is the inclusive cross section of $e^- e^+ \to t \bar t$ with unpolarised $e^+ e^-$ beams, 
$\sigma_x( e^- e^+ [\rin^{(a,b)}] \to t \bar t )$ is the effective  
cross section after the kinematic selection, $x$, of the $t \bar t$ momenta, for the initial polarisation state $\rin^{(a,b)}$ and $\rho_x^{(a,b)}$ is the $t \bar t$ spin density matrix.
The method of experimental reconstruction of the effective $t\bar{t}$ spin density matrix ({\it quantum state tomography}) is studied and outlined in \cite{Afik:2020onf,Ashby-Pickering:2022umy}, 
and has recently been applied in $pp$ collisions at the LHC by CMS~\cite{CMS:2024zkc}. 

To facilitate the comparison with the theoretical prediction \eqref{IxTheo} for the Choi matrix we shall use formula \eqref{Choi_exp}. 
It yields the blocks of the experimentally reconstructed Choi matrix:
\begin{align}
{\cal I}_x (\ket{IJ}\bra{KL}) = \sum_{a,b=1}^4 
A^{(I,K)}_{a}
\,
B^{(J,L)}_{b}\,\cdot
\varrho_x^{(a,b)},
\end{align} 
In particular, the diagonal block is reconstructed as
\begin{align}
{\cal I}_x (\ket{++}\bra{++}) & = \frac{1}{4 q \bar{q}} \Big[ (1+q)(1+\bar{q}) \varrho_x^{(+,+)} - (1+q)(1-\bar{q}) \varrho_x^{(+,-)}  \notag\\
& \qquad\qquad  - (1-q)(1+\bar{q}) \varrho_x^{(-,+)} + (1-q)(1-\bar{q}) \varrho_x^{(-,-)} \Big],
\end{align}
which only uses the data from the longitudinally polarised runs. On the other hand, the off-diagonal block reconstruction requires the data from all sixteen runs:
\begin{align}
{\cal I}_x (\ket{++}\bra{--}) & = \frac{1}{2 q \bar{q}} \Big\{ i \big[ \varrho_x^{(+,+)} +\varrho_x^{(+,-)} +\varrho_x^{(-,+)} +\varrho_x^{(-,-)} \big] + \notag \\
& \qquad\quad - (1+i) \big[ \varrho_x^{(+,\vec{x})} +\varrho_x^{(-,\vec{x})} +\varrho_x^{(\vec{x},+)} +\varrho_x^{(\vec{x},-)} \big] + \notag \\
& \qquad\quad + (1-i) \big[ \varrho_x^{(+,\vec{y})} +\varrho_x^{(-,\vec{y})} +\varrho_x^{(\vec{y},+)} +\varrho_x^{(\vec{y},-)} \big] + \notag \\
& \qquad\quad + 2 \big[\varrho_x^{(\vec{x},\vec{x})} + i \varrho_x^{(\vec{x},\vec{y})}  + i \varrho_x^{(\vec{y},\vec{x})} -  \varrho_x^{(\vec{y},\vec{y})} \big] \Big\}.  
\end{align}

The tree-level Standard Model prediction indicates that twelve $4 \times 4$ blocks of the Choi matrix \eqref{Choi_eett} vanish. 
Experimentally, this would be seen as a cancellation among the $\varrho_x^{(a,b)}$ outcomes.
For instance,
\begin{align}
{\cal I}_x (\ket{+-}\bra{+-}) & = \frac{1}{4 q \bar{q}} \Big[ -(1+q)(1-\bar{q}) \varrho_x^{(+,+)} + (1+q)(1+\bar{q}) \varrho_x^{(+,-)} + \notag\\
& \qquad\qquad  + (1-q)(1-\bar{q}) \varrho_x^{(-,+)} - (1-q)(1+\bar{q}) \varrho_x^{(-,-)} \Big]
\end{align} 
should vanish as a 4 $\times$ 4 matrix.

\section{Experimental implementation}
\label{sec:practicalities}

The procedure we have outlined relies on making repeated quantum-state-tomography measurements of the output for a set of differently prepared input states. 
In our example the different input states correspond to different beam spin polarisations, which leads to a need to control and manipulate those beam polarisations. More generally, a variety of polarised beam sources could be employed. 

\subsection{Polarised fermion beams}

Circular accelerators can make use of the Sokolov–Ternov effect \cite{Sokolov:1963zn}, the natural transverse self-polarization of relativistic electrons or positrons through spin-flip synchrotron radiation. 
In an ideal planar ring the beam could be polarised 
up to 92.4\% \cite{Buon:1993rtf}. 
For the different configuration of a linear collider, a longitudinally polarised electron beam can be produced by a laser-driven photo injector, where circularly polarized photons illuminate a photocathode \cite{doi:10.1142/S201019451660003X}. 

As the beam approaches the interaction point, the polarisation can be rotated in arbitrary directions around the Bloch sphere through the use of specialised spin rotator magnets.
In practice, polarisations of electron beams of $70-80\%$ have been obtained for multi-GeV electron and positron beams~\cite{Buon:1993rtf,ParticleDataGroup:2024cfk}. 
At the HERA $ep$ collider 30-45\% longitudinal polarisation was obtained at the interaction points~\cite{sobloher2012polarisationherareanalysis,ParticleDataGroup:2024cfk}. The linear collider SLC operated 45.6\,GeV beams with typically 75\% electron-beam polarisation \cite{SLD:2000leq}. 

Beam polarisation is also possible with protons and ions. 
The COMPASS fixed-target experiment \cite{COMPASS:2007rjf} directs high-energy muon and hadron beams at a large polarized target inside a superconducting solenoid. 
The RHIC collider has achieved 60\% polarised proton beams at beam energies of 255\,GeV \cite{Raparia:2024npt}. 
The EIC plans to have highly polarized $\approx$70\% polarised beams of both electrons and nucleons, and is developing special magnets and operation techniques to preserve polarisation through acceleration to collisions. 

In considering experiments with polarised hadrons, one might wish to consider the initially prepared `state' as being that of the constituent quarks inside these hadrons. 
In this case, one would need to account for the fact that the degree of polarisation of those quarks is different than, and generally smaller than that of the hadrons \cite{Bass:2004xa}. 
In some cases, particularly when dealing with QCD processes at lower energies, it would be more appropriate to describe the initial state in terms of the hadrons themselves. 
Both perspectives are valid as mechanisms for performing the types of quantum tests we are proposing, and there would be value in investigating the relative merits of each.

Looking towards the future, proposed $e^+e^-$ linear colliders have been designed to deliver highly polarised beams. At the proposed international linear collider (ILC) the specification is for electron polarisation of 80\%, with positron polarisation of 30\% in the baseline, and an upgrade option for positron polarisation of 60\% \cite{Adolphsen:2013kya}. The CLIC linear collider proposes 80\% $e^-$ beam polarisation; it has no positron polarisation specified in the baseline design, but that might be considered as an upgrade. It is more difficult to maintain beam polarisation at high luminosity circular colliders, but it is possible to use resonant depolarisation of beams to precisely measure the beam energy, as was done at LEP \cite{Arnaudon:1994zq}. Studies for the proposed FCC-ee collider propose a goal of 10\% beam polarisation at 45 and 80\,GeV beam energies for measurements at the $Z$ boson pole and the $W^\pm$ pair threshold \cite{Blondel:2019jmp}. 

\subsection{Polarised boson beams}

While we have concentrated on polarised beams of charged fermions, a developing possibility is that of highly polarised beams of multi-GeV photons. Such beams are thought to be achievable through the interaction of a highly intense laser pulse with a spin-polarized counter-propagating ultrarelativistic electron beam. 
Polarisations of up to 95\% have been predicted in recent calculations \cite{Gamma_pol1,Gamma_pol2}.

A further class of polarised `initial state' beams is that of $W$ and $Z$ bosons emitted from high-energy protons, which are often considered in vector-boson-scattering or vector-boson-fusion processes at hadron colliders. While it is not generally possible to directly manipulate the spin of these spin-1 bosons, they are generally produced in association with jets from the quarks from which they are emitted. Different spin configurations for these `initial state' bosons can be post-selected by selecting particular kinematical configurations of the jets. Then the polarisation of the bosons can be calculated in a manner similar to that described in Ref.\ \cite{Cheng:2024rxi}. 
Thus one could perform this `kinematic' quantum state tomography of the initial state, combined with the decay quantum state tomography on the final state, offering the possibility of calculating the Choi matrix for the entire scattering process.

\subsection{Practicalities}

While the polarisation of charged-fermion beams we have described can be rotated to arbitrary directions through spin rotator magnets, we note that it would be necessary to plan a dedicated running period for each selected combination of beam polarisations. Each of these periods would require sufficient integrated luminosity to be collected to allow quantum state tomography on the final state process. 
For full quantum process tomography of the full Choi matrix described in section \ref{sec:tomography} one requires 16 different sets of beam conditions. 

In Ref.\ \cite{Maltoni:2024csn}, a rough estimation of the statistical uncertainty on the $t \bar t$ spin density matrix reconstruction at an $e^+ e^-$ collider has been provided, assuming similar measurements in \cite{ATLAS:2023fsd, CMS:2019nrx}.
Expressing the $t \bar t$ spin density matrix as
\be
\rho_{t \bar t} = \frac{1}{4} \left[ 1 
+ B_i \sigma_i \otimes {\mathbb 1} 
+ \overline B_i {\mathbb 1} \otimes \sigma_i 
+ C_{ij} \sigma_i \otimes \sigma_j 
\right] ,~~~(i,j=1,2,3)
\ee
their result implies that the uncertainty on the diagonal elements of $C$ is $\sim 3.6 / \sqrt{ \epsilon N}$
and $\sim 1.8 / \sqrt{ \epsilon N}$ for the dilepton and  lepton+jet final states, respectively, where $N = {\cal L} \cdot \sigma$ is the total number of $t \bar t$ events for a given luminosity, $\cal L$, and the $e^+ e^- \to t \bar t$ cross section, $\sigma$, and $\epsilon \sim 0.4$ is the global reconstruction efficiency.   
The uncertainty on the non-zero off-diagonal elements of $C$ is estimated to be $\sim 14.3 / \sqrt{\epsilon N}$ and $\sim 7.0 / \sqrt{\epsilon N}$ for the dilepton and lepton+jet final states, respectively.
The measurement of $B_i$ ($\overline B_i$) is essentially a polarisation measurement of $t$ ($\bar t$), which is more straightforward than the spin correlation measurements for $C_{ij}$.  
We therefore expect better accuracies for $B_i$ ($\overline B_i$) for given collision energy and luminosity.  
\begin{figure}[t!]
    \centering
    \includegraphics[width=0.5\linewidth]{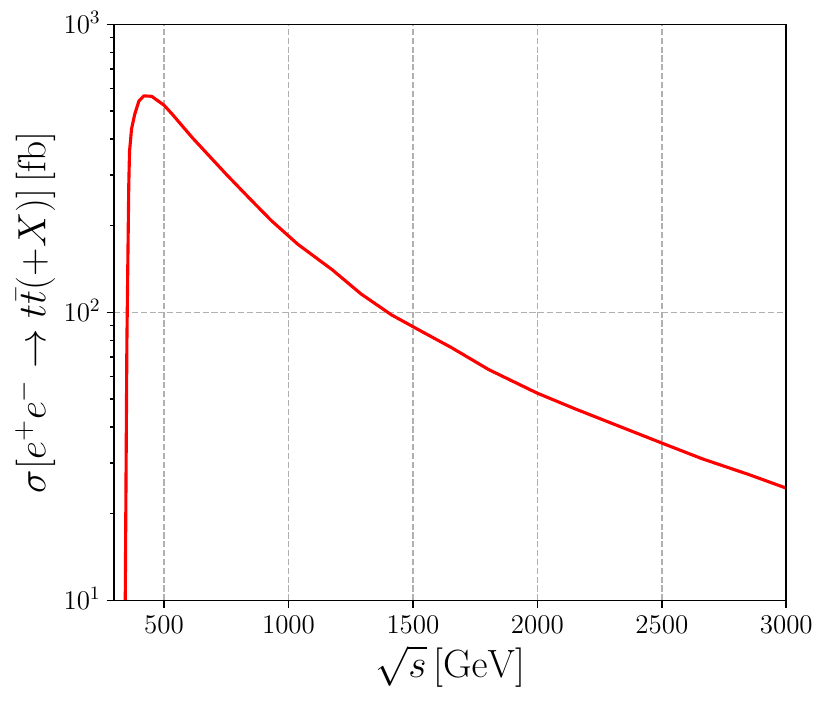}
    \caption{\small The $t \bar t$ production cross section at an unpolarised $e^+ e^-$ collider as a function of the collision energy $\sqrt{s}$. 
    The values are taken from \cite{CLICdp:2018cto}.}
    \label{fig:xsec}
\end{figure}
As a reference, we show in Fig.\ \ref{fig:xsec} the $t \bar t$ production cross section at an unpolarised $e^+ e^-$ collider estimated in \cite{CLICdp:2018cto} as a function of the beam collision energy.
Around $\sqrt{s} = 400$ GeV, the cross section peaks to give $\sigma \sim 580$ fb, while at $\sqrt{s} = 1000$ GeV, it is $200$ fb.  
Demanding $\sim 10\,\%$ accuracy for the $C_{ij}$ ($i \neq j$) measurement, the required luminosities are estimated to be $\sim 20$ fb$^{-1}$ and $\sim 60$ fb$^{-1}$ for $\sqrt{s} \sim 400$ and 1000 GeV, respectively.    

In the proposed method, one must repeat such a measurement 16 times with different polarisation settings. While the polarisation of the beams need not be very high, a low degree of polarisation, as well as its uncertainty, will influence the inverse matrices in Eq. \eqref{AB} and hence increase the statistical uncertainty of the estimated Choi matrix elements.

The practicality of implementing the 16 runs for different polarisations and the determination of the required luminosity are an optimisation problem that will depend on the particular collider, the polarisation that can be achieved, the acceptance of the detector, and the final state selected. While such studies go beyond the scope of this initial proposal, 
we note that the rich landscape of polarised beams and final states makes this a fruitful area for future research.

\section{Conclusion}
\label{sec:conclusion}

We have outlined a procedure by which the full Choi matrix of a high-energy scattering process can be determined and tested using beams with differing polarisations and performing quantum state tomography with each setting. 

The procedure treats the scattering process as a quantum instrument and opens the possibility of performing a variety of tests on novel systems. Applications include tests of the Standard Model, searches for new particles and fields, and foundational tests of quantum theory. 

While we have illustrated the technique -- quantum process tomography -- using the theoretically clean process $e^- e^+ \to t \bar t$, the method is rather general. It applies to any process at any collider for which initial state preparation is possible and for which quantum state tomography is possible in the final state. 
Initial states could include polarised electrons or hadrons, or potentially even photons or weak bosons. 
Final states of interest include top quarks and bottom hadrons, $\Lambda$ baryons, $\tau$ leptons, $W^\pm$ bosons, $Z^0$ bosons and various spin-1 mesons. 

The ability to marry the various different production mechanisms with different final states generates a large number of possible processes which can be investigated in future work and at future accelerators and colliders.

\section*{Acknowledgements}
AJB is grateful to Phil Burrows and Jenny List for helpful exchanges regarding the capabilities of future colliders. 
CA is particularly grateful to Malcolm Fairbairn for helpful comments and discussion. 
The work of KS was performed in part at the workshop on {\it Effective Theories for Nonperturbative Physics} at the Mainz Institute for Theoretical Physics
(MITP) of the DFG Cluster of Excellence PRISMA+ (Project ID
390831469).
We thank the organizers of two workshops: the first \textit{Quantum Observables for Collider Physics}, November 2023, funded by the  Galileo Galilei Institute for Theoretical physics of the \textit{Istituto Nazionale di Fisica Nucleare}, and the second \textit{Quantum Tests in Collider Physics}, October 2024, supported by the Institute for Particle Physics Phenomenology, Durham, and Merton College, Oxford and for hosting several of the authors during the preparation of this manuscript.
AJB is funded through STFC grants ST/R002444/1 and ST/S000933/1, by the University of Oxford, and by Merton College, Oxford. 
CA is funded through a King’s College London NMES studentship. ME was supported by the National Science Centre in Poland under the research grant Sonata BIS (2023/50/E/ST2/00472). PH acknowledges support by the National Science Centre in Poland under the research grant Maestro (2021/42/A/ST2/0035). 

\section*{Supporting Information}

\appendix 

\section{Choi matrix reconstruction} \label{app:tomo}

In this Appendix we show how to reconstruct the Choi matrix \eqref{ChoiE} of a quantum channel from experimentally gathered data $\{\rin^k,\rout^k\}_{k=1}^{\din^2}$.

The initial states are chosen so that they form a basis of the space $\S(\Hin)$. We then introduce the set of operators $\rin^{k,\perp}$ that is bi-orthonormal to $\rin^k$ in the following sense:
\begin{align}
\Tr(\rin^k\rin^{l,\perp})= \delta_{kl}.
\end{align}
On the side, we note that the bi-orthonormality can also be understood in terms of the inner product $\langle \langle A | B \rangle \rangle = \Tr(A^{\dagger} B)$ between any two operators $A, B \in {\cal B}({\cal H}_{in})$.
By a standard fact from linear algebra, such a bi-orthonormal set 
exists and forms a (typically non-orthogonal) basis, since the states $\rin^l$ form a basis of ${\cal B}({\cal H}_{in})$.

Given the standard representation of the input matrices $\rin^{k}
=\sum_{i,j=1}^{\din} R_{ij}^{k} |i \rangle \langle j|$ one can find explicitly the standard matrix representation of the bi-orthonormal basis elements (not positive semi-definite in general) 
\begin{align}
\rin^{k,\perp}
=\sum_{i,j=1}^{\din} R_{ij}^{k,\perp} |i \rangle \langle j|
\label{Representation}
\end{align} 
by solving, for any fixed $k$, the following set of $\din^{4}$ equations:
\begin{align}
\sum_{i,j=1}^{\din} R_{ji}^{k}R_{ij}^{l,\perp} = \delta_{kl}, \quad \text{ for } \quad k,l=1, \ldots, \din^{2}.
\label{Solution}
\end{align}
Each solution corresponds to the matrix representation  of a particular bi-orthonormal basis element:
\begin{align}\label{Inversion1}
R_{ij}^{k,\perp}
=\langle i|\rin^{k,\perp} |j \rangle. 
\end{align}

Then, the Choi matrix corresponding to the channel $\E$ can be expressed as 
\begin{align}
\widetilde{\E} = \frac{1}{\din} \sum_{k = 1}^{\din^2} (\rin^{k,\perp})^{T}  \otimes \rout^k.
\label{BiorthRepresentation}
\end{align}
Before proving the above equality, let us recall the well-known fact that {\it any} operator $\widetilde{\F} \in \B(\Hin \otimes \Hout)$ uniquely defines some linear map $\F: \B(\Hin) \to  \B(\Hout)$ defined, via the Choi--Jamio{\l}kowski isomorphism as
\be
\widetilde{\F} \,=\, \frac{1}{\din} \sum_{i,j} | i \ketbra j |  \otimes {\cal F}(| i \ketbra j |) \,.
\label{Rep1}
\ee
Furthermore, on can check that the action of the map $\F$  on any state $\rho$ can be expressed as
\begin{align}
\F(\rho)=d_{\rm in}\Tr_{{\cal H}_{\rm in}}[(\rho^{T} \otimes 1 ) \widetilde{\F}]
\label{Rep2}
\end{align}
 Now observe that (\ref{Rep1})  implies that the RHS of  
 (\ref{BiorthRepresentation}) uniquely defines some map $\cal G$. 
 In other words, according to the Choi--Jamio{\l}kowski isomorphism, the following operator
$\widetilde{\cal G}:=\frac{1}{\din} \sum_{k = 1}^{\din^2} (\rin^{k,\perp})^{T}  \otimes \rout^k$ 
can always be uniquely represented as 
$\widetilde{\cal G}=
\frac{1}{\din} \sum_{i,j} | i \ketbra j |  \otimes {\cal G}(| i \ketbra j |)$ for some, yet unknown, but uniquely defined, map ${\cal G}:\B(\Hin) \to  \B(\Hout)$.
To complete the proof of (\ref{BiorthRepresentation}) it is enough to show that the two maps satisfy ${\cal  G} = {\cal E}$, since if the maps are the same, then, by uniqueness of (\ref{Rep1}) their Choi-Jamo{\l}kowski representations $\widetilde{\cal  G}$ are $\widetilde{\cal  E}$ the same. 
Using (\ref{Rep2}) we can calculate immediately the action of ${\cal  G}$ 
on any basis elements $\rin^{k}$ which gives ${\cal  G}(\rin^{k})=\rout^{k} :={\cal  E}(\rin^{k})$. 
By linearity, since both maps act identically on all the basis elements, they are identical.  
This completes the proof of \eqref{BiorthRepresentation}.

We note that $\rout^k$ may not form a basis (which happens, for example, in the case of so-called dumping channels, where the channel is not invertible). 
However, this does not cause any problems in the present analysis.

We also wish to understand how to represent the Choi matrix in the standard basis. To do this, let us 
notice that the equations~\eqref{Solution}
show how to transform the input density matrix basis $\{ \rin^{k} \}$ into bi-orthonormal matrix basis $\{ \rin^{k,\perp} \}$ (which is in some sense canonical, see \eqref{Identity} below). However we may also map the input basis  $\{ \rin^{k} \}$ into the standard
matrix basis $\{ |i \rangle \langle j|\}$
and the corresponding transformation is tightly related to the bi-orthonormal elements. 
In fact, the set of equations \eqref{Solution} is equivalent to finding the inverse of the $\din^2 \times \din^2$ matrix $[Q_{kn}]$, where $Q_{kn} = R^k_{ij}$ with $n = n(i,j) = (i-1) \din + j$, for $i,j \in \{1,\ldots,\din\}$. The matrix $Q$ is non-singular, as it transform the basis $\{ | i \ketbra j | \}_{i,j=1}^{\din}$ into the basis $\{\rin^k\}_{k=1}^{\din^2}$ of the space $\B(\Hin)$:
\begin{align}
  \rin^{k} =\sum_{i,j=1}^{\din} R_{ij}^{k} |i \rangle \langle j|  = \sum_{n=1}^{\din^2} Q_{k,n(i,j)} |i \rangle \langle j| \quad \Rightarrow \quad |i \rangle \langle j| = \sum_{n=1}^{\din^2} (Q^{-1})_{n(i,j),k} \rin^k.
\end{align}
Note that $Q^{-1})_{n(i,j),k} = X^{(i,j)_k}$ with $X$'s defined in formula \eqref{TheNewX} the main text.
Observe that the map
\begin{align}
{\cal F}_{\rm Id}(\cdot)= \sum_{k=1}^{\din^{2}} | \rin^{k} \rangle \rangle \langle \langle  \rin^{k,\perp} | (\cdot) : = 
\sum_{k=1}^{\din^{2}} \rin^{k} {\rm Tr}[\rin^{k,\perp} \cdot]
\label{Identity}
\end{align}
is the identity map since it maps all the basis elements $\rin^{k}$ into themselves. Hence,
\begin{align}\label{bi_inv}
|i\rangle\langle j| = {\cal F}_{\rm Id}(|i\rangle\langle j|)=
\sum_{k=1}^{\din^{2}} \rin^{k} {\rm Tr}[\rin^{k,\perp} |i\rangle \langle j|]=
\sum_{k=1}^{\din^{2}} \langle j| \rin^{k,\perp} | i \rangle \rin^{k}.
\end{align}
and we have 
\begin{align}
R^{k,\perp}_{ij} =  (Q^{-1})_{n(j,i),k} =  \big((Q^{-1})_{n(i,j),k} \big)^*. 
\end{align}
Consequently,
\begin{align}
\widetilde{\E} := \frac{1}{\din}
\sum_{i,j=1}^{\din} \ket{i}\bra{j} \otimes \E(\ket{i}\bra{j}) =
\frac{1}{\din}
\sum_{i,j=1}^{\din} \ket{i}\bra{j} \otimes \sum_{k = 1}^{\din^2} R^{k,\perp}_{ji}  \rout^k.
\end{align}

Let us now consider the case when the Hilbert space $\Hin$ is a tensor product, $\Hin = \H_1 \otimes \H_2$ with $d_i = \dim \H_i$. If $\{\rho_{\ini,1}^a\}_{a=1}^{d_1^2}$ and $\{\rho_{\ini,2}^b\}_{b=1}^{d_2^2}$ are the bases of $\S(\H_1)$ and $\S(\H_2)$, respectively, then 
\begin{align}\label{rin_tens}
    \rin^k = \rho_{\ini,1}^a \otimes \rho_{\ini,2}^b, \quad \text{ with } \quad k = (a-1) d_2^2 + b 
\end{align}
is a basis of $\S(\H_1) \otimes S(\H_2)$. 
For the bi-orthogonal elements, we can simply take
\begin{align}
    \rin^{k,\perp} = \rho_{\ini,1}^{a,\perp} \otimes \rho_{\ini,2}^{b,\perp}.
\end{align}
Indeed,
\begin{align}
    \Tr \big( \rho_{\ini,1}^a \otimes \rho_{\ini,2}^b \big) \big( \rho_{\ini,1}^{c,\perp} \otimes \rho_{\ini,2}^{d,\perp} \big) = \big( \Tr_{\H_1} \rho_{\ini,1}^a \rho_{\ini,1}^{c,\perp} \big) \cdot \big( \Tr_{\H_2} \rho_{\ini,2}^b \rho_{\ini,2}^{d,\perp} \big) = \delta_{ac} \delta_{bd}.
\end{align}
Hence, to reconstruct the Choi matrix for a channel with the initial Hilbert space being a tensor product, it is sufficient to prepare the initial states as tensor product states and find the corresponding bi-orthogonal elements independently for each of the two components. 
This obviously extends to any higher tensor product states of $\Hin$.  

The above procedure can also be applied to quantum instruments if we replace ${\cal E}$ with ${\cal I}_x$
and $\widetilde{\cal E}$ with $\widetilde{\cal I}_x$ with one significant, but simple change. 
Namely, we should replace the output states $\rout^{k}$ by the unnormalised  $\varrho_{x}^{k}=\widetilde{\cal I}_x(\rin^{k})$ (cf. 
(\ref{Unnormalised-State})).
This is related directly to the fact 
that a quantum instrument 
has effectively two outcomes:
 (i) the probability $p_{k}=\Tr(\widetilde{\cal I}_x(\rin^{k}))=\Tr(\varrho_x^{k})$ that an output within the range $x$ was produced and (ii) the output state $\rho_x^{k}=\varrho_{x}^{k}/p_{k}$
produced with this probability (defined whenever $p^{k}>0$).

In a typical quantum process tomography scenario, one tries to arrange for the input states to be pure \cite{PTomography1},
\begin{align}
\rin^k = |v_k \ketbra v_k|, \quad \text{ for } \quad \ket{v_k} \in \Hin.
\end{align}
This guarantees that we have maximal knowledge about the relevant quantum degree of freedom of the input system.

A particularly simple class of pure input states is determined by so-called symmetric informationally complete positive operator-valued measures (SIC-POVMs) \cite{SICPOVM1,SICPOVM2}. The latter are $d^{2}$ pure states $\Pi^{k}=|\Psi^{k} \rangle \langle \Psi^{k}|$ defined on $d$-dimensional Hilbert space, satisfying 
the following relation:
\begin{align}
\Tr(\Pi^{k}\Pi^{l})=\frac{d \delta_{kl} + 1}{d+1}
\end{align}
In this case, given the explicit form of $\Pi^{k}$, finding bi-orthonormal elements $\Pi^{k,\perp}$ is immediate to calculate and gives:
\begin{align}
\Pi^{k,\perp}=\frac{1}{d}\big( (d+1)\Pi^{k} - {\bf 1}\big)
\end{align}
Setting $\rin^{k}=\Pi^{k}$, $\rin^{k,\perp}=\Pi^{k,\perp}$ yields
the matrix elements \eqref{Inversion1} needed to compute the Choi matrix \eqref{BiorthRepresentation}.

This calculation can be quickly generalised to the case of noisy inputs 
$\rin^{k}=q \Pi^{k} + \tfrac{1-q}{d} {\bf 1}$, for any $q \in (0,1]$. 
Let us illustrate this for a single-qubit channel. Consider the regular tetrahedron defined by 4 unit vectors in $\mathbb{R}^{3}$ satisfying 
\begin{align}
\hat{n}_{k}\cdot \hat{n}_{l}=
\tfrac{4}{3}\delta_{kl} - \tfrac{1}{3}.
\end{align}
Probably the most elegant example of such four vectors are 
\begin{align}
  \hat{n}_{1}=[\sqrt{\tfrac{1}{3}}, \sqrt{\tfrac{2}{3}},0], \quad \hat{n}_{2}=[\sqrt{\tfrac{1}{3}}, - \sqrt{\tfrac{2}{3}},0], \quad
\hat{n}_{3}=[-\sqrt{\tfrac{1}{3}},0, -\sqrt{\tfrac{2}{3}}], \quad \hat{n}_{4}=[-\sqrt{\tfrac{1}{3}},0, \sqrt{\tfrac{2}{3}}],
\end{align}
which form a regular tetrahedron in the Bloch sphere \cite{SICPOVM2}.
It can be easily seen that 
the four rank-1 projectors  $\Pi^{k} =\frac{1}{2}({\bf 1} +  \hat{n}_{k} \cdot \vec{\sigma})$ corresponds to a SIC-POVM, i.e.\ they satisfy $\Tr(\Pi^{k}
\Pi^{l})=\frac{2 \delta_{kl} + 1}{3}$. 
Suppose that our input states correspond to their noisy variants, i.e. 
\begin{align}
\rin^{k} =\tfrac{1}{2}({\bf 1} + q \hat{n}_{k} \cdot \vec{\sigma}), \qquad k=1,2,3,4.
\label{qubit-input}
\end{align}
Because of the symmetry, it is immediate to find explicitly the bi-orthonormal elements 
\begin{align}
\rin^{k,\perp} =\tfrac{1}{4}\big({\bf 1} + \tfrac{3}{q} \hat{n}_{k} \cdot \vec{\sigma}\big).
\end{align}

Quite remarkably, if we make the elements of the SIC-POVM noisy in a way which corresponds to shrinking the Bloch vectors by different factors $q_k$ in the formula 
(\ref{qubit-input}) 
(of which only one can be equal to zero), the corresponding states still form a legitimate basis of the space $\S(\mathbb{C}^2)$.

In the example of polarised $e^-e^+ \to t \bar{t}$ scattering considered in section \ref{sec:tomography}, the input states $\rin^k$ correspond to two uncorrelated polarisation states of the beams of colliding electrons and positrons. Let us first focus on the $e^-$ beam alone. As explained in section \ref{sec:tomography}, we need four different initial polarisations, among which two can be conveniently chosen to be along the beam ($z$-axis):
\begin{align}\label{rin_z1}
    \rho_{\ini,e}^1 & = q_{e} \ket{+}\bra{+} +\tfrac{1}{2} (1 - q_e) {\bf 1} = \tfrac{1}{2} (1+q_{e}) \ket{+}\bra{+} +\tfrac{1}{2} (1 - q_e) \ket{-}\bra{-},\\
    \rho_{\ini,e}^2 & = q_{e} \ket{-}\bra{-} +\tfrac{1}{2} (1 - q_e) {\bf 1} = \tfrac{1}{2} (1+q_{e}) \ket{-}\bra{-} +\tfrac{1}{2} (1 - q_e) \ket{+}\bra{+}, \label{rin_z2}
\end{align}
where $q_e \in [0,1]$ is the degree of polarization. The other two initial polarisations need to have a transverse component:
\begin{align}
    \rho_{\ini,e}^3 & = q_{e}(\m) \ket{\m}\bra{\m} +\tfrac{1}{2} (1 - q_{e}(\m) ) {\bf 1} = \tfrac{1}{2} (1+q_{e}(\m) ) \ket{\m}\bra{\m} +\tfrac{1}{2} (1 - q_{e}(\m)) \ket{-\m}\bra{-\m},
    \nonumber \\
    \rho_{\ini,e}^4 & = q_{e}(\n)  \ket{\n}\bra{\n} +\tfrac{1}{2} (1 - q_{e}(\n) ) {\bf 1} = \tfrac{1}{2} (1+q_{e}(\n) ) \ket{\n}\bra{\n} +\tfrac{1}{2} (1 - q_{e}(\n) ) \ket{-\n}\bra{-\n},
\end{align}
where $\m, \n$ are two, not aligned, polarisation directions. We assume that, in general, the degree of the polarisation $q_{e}(\m)$ may depend on the direction. With the notation \eqref{pure_pol} we can write down explicitly the transformation matrix from the standard basis $\{ | i \ketbra j | \}_{i,j=\pm}$ of $\S(\Hin)$ to the one spanned by $\{\rho_{\ini,e}^k\}_{k=1}^4$,
\begin{align}
Q =    \left(
\begin{array}{cccc}
 \frac{1+ q_e}{2} & 0 & 0 & \frac{1-q_e}{2} \\
 \frac{1-q_e}{2} & 0 & 0 & \frac{1+q_e}{2} \\
 |\alpha|^2  q_e(\m) + \frac{1}{2} (1-q_e(\m)) & \alpha   \beta ^* q_e(\m) & \beta  \alpha ^* q_e(\m)
   & |\beta|^2  q_e(\m) + \frac{1}{2} (1-q_e(\m)) \\
 |\gamma|^2  q_e(\n) + \frac{1}{2} (1-q_e(\n)) & \gamma  \delta ^* q_e(\n) & \delta  \gamma q_e(\n)
   ^* & |\delta|^2  q_e(\n) + \frac{1}{2} (1-q_e(\n)) \\
\end{array}
\right).
\end{align}
In order to find the bi-orthogonal elements $\rin^{l,\perp}$, needed for the reconstruction of the Choi matrix \eqref{BiorthRepresentation}, we simply need to find the inverse of the matrix $Q$. 
For an illustration, let us take $\m = \vec{x}$ and $\n = \vec{y}$, that is $\alpha = \beta = \gamma = 1/\sqrt{2}$, $\delta = i/\sqrt{2}$, and set $q_e(\m) = q_e(\n) = q_e^T$. 
Then, the corresponding bi-orthogonal elements read
\begin{align}
 \rho_{\ini,e}^{1,\perp} = \begin{pmatrix}
     \frac{1+q_e}{2 q_e} & \frac{i-1}{2 q_e^T} \\ -\frac{i+1}{2 q_e^T} & \frac{q_e-1}{2 q_e}
     \end{pmatrix},
&&
   \rho_{\ini,e}^{2,\perp} = \begin{pmatrix}
     \frac{q_e-1}{2 q_e} & \frac{i-1}{2 q_e^T} \\ -\frac{i+1}{2 q_e^T} & \frac{1+q_e}{2 q_e}
     \end{pmatrix},
&&
   \rho_{\ini,e}^{3,\perp} = \frac{1}{q_e^T} \begin{pmatrix}
     0 & 1 \\ 1 & 0
     \end{pmatrix},
&&
   \rho_{\ini,e}^{4,\perp} = \frac{1}{q_e^T} \begin{pmatrix}
     0 & -i \\ i & 0
     \end{pmatrix}. 
\end{align}
The same calculation can be done for the positron beam --- we can keep the same directions of polarisations, but allow for different degrees of polarisation $q_{\bar{e}}(\m)$. 
In fact, in lepton collider experiments the positron beam has a smaller degree of polarisation than the electron beam. 
Then, we construct the bi-orthogonal elements for the tensor product subspace of $S(\H_1 \otimes \H_2)$, as explained around Eq.\ \eqref{rin_tens}. 
We thus have all elements needed to reconstruct the Choi matrix \eqref{Choi_exp} from the gathered experimental data.

Finally, let us note that one of the initial states, say \eqref{rin_z1}, can be taken to be maximally mixed, $\rho_{\ini,e}^1 = \tfrac{1}{2} {\bf 1}$. However, we do need to have a non-zero polarisation degree in the second state \eqref{rin_z2}. This means that we essentially need to polarise (imperfectly) each of the initial beams in three directions: 1 along the beam and 2 different transversal ones + one unpolarised run. 
For two beams we need all 16 combinations: 9 with both beams polarised, 6 with one beam polarised and the second not and 1 with both unpolarised beams. Alternatively, and equivalently, we can design the 16 runs with the set of polarisations listed at the end of Section~\ref{sec:tomography} or the
16 runs of the $4 \times 4$ products of a single qubit SIC-POVM polarisations given by Eq.\ \eqref{qubit-input}.

\section{Rescaling of $\varrho_x'$}
\label{app}

\subsection{The derivation}

In this appendix, we demonstrate how the outcome of the quantum channel in collider physics, $\varrho_x'$ in Eq.\ \eqref{QI}, becomes infinitesimally small since it is proportional to a dimensionful factor which vanishes in the limit of infinite time and volume.
The procedure to rescale $\varrho_x'$ to obtain $\varrho_x$ in Eq.\ \eqref{varrho_form} is also shown.

In our demonstration, we note that the spin/flavour part of $\varrho_x'$ in Eq.\ \eqref{QI} is linear and well-behaved.
It is, therefore, sufficient to focus on the momentum part. 
For the scattering $\alpha \beta \to \gamma \delta$, we thus analyse 
\be
\varrho_x' = 
\int_x d \Pi_{\gamma \delta}  \,
 \bra{p_f} 
 S | \tildepin \ketbra \tildepin | S^\dagger 
 \ket{p_f}.
 \label{QI2}
\ee
and show that this can be rescaled to 
\be
\varrho_x =
\frac{1}{\sigma_{\cal N}}
\cdot 
\frac{1}{2 s}
\int_x d \Pi_{\rm LIPS}   
 {\cal M}^{\pin}_{p_f} 
 \left[ {\cal M}^{\pin}_{p_f} \right]^* \,.
 \label{varrho2}
\ee

First, we construct the initial state $\ket{\tildepin}$ with the proper normalisation, $\braket{ \tildepin | \tildepin } = 1$, out of the Lorentz-covariantly normalised kets
\be
\braket{k_1,k_2|k_1',k_2'} = (2 \pi)^6 4 E_{k_1} E_{k_2} \delta^3( {\bf k}_1 - {\bf k}_1' )
\delta^3( {\bf k}_2 - {\bf k}_2' ).
\ee
We assume that the initial particles $\alpha$ and $\beta$ have momentum distributions that sharply peak at ${\bf p}_\alpha$ and ${\bf p}_\beta$, respectively. 
We construct such a state with the two momentum-space wave functions, $\phi_\alpha( k_1)$ and $\phi_\beta( k_2)$, as
\be
\ket{\tildepin} = 
\int \frac{d^3 k_1}{(2 \pi)^3 } \frac{1}{\sqrt{2 E_{k_1}}} \phi_\alpha( k_1 ) 
\frac{d^3 k_2}{(2 \pi)^3 } \frac{1}{\sqrt{2 E_{k_2}}} \phi_\beta( k_2 )
\ket{ k_1, k_2 }
\label{ptilde}
\ee
By assumption, the momentum wave functions $\phi_{X_i}( k_i)$ sharply peaks at ${\bf k}_i = {\bf p}_{X_i}$ for $i=1,2$ and $(X_1, X_2) = (\alpha,\beta)$ and are normalised as
\be
\int \frac{d^3 k_i}{(2 \pi)^3} \left| \phi_{X_i}(k_i) \right|^2 = 1
\ee
to imply $\braket{\tildepin|\tildepin} = 1$.
One can, therefore, formally write the wave function as
\be
\phi_{X_i}(k_i) = \frac{(2 \pi)^3 \delta^3({\bf k}_i - {\bf p}_{X_i} ) }{ \sqrt{V} } 
\label{pwave}
\ee
with the volume of all space 
\be
V = \left[ \int d^3 x \, e^{i {\bf p} \cdot {\bf x}}  \right]_{ {\bf p} = {\bf 0} } = (2 \pi)^3 \delta^3( {\bf 0} )\,.
\ee

In the expression \eqref{QI2}, 
$\ket{p_f} = \ket{p_\gamma, p_\delta}$ is Lorentz-covariantly normalised.
Writing $S = 1 + i {\cal T}$,
the transfer matrix element $\bra{p_f} S \ket{\tildepin} = \bra{p_f} i {\cal T} \ket{\tildepin}$ as
\bea
\bra{p_f} i {\cal T} \ket{\tildepin}
&=&
\int \frac{d^3 k_1 d^3 k_2}{(2 \pi)^6}
\frac{1}{\sqrt{4 E_1 E_2}}
\frac{ (2 \pi)^6 \delta^3( {\bf k}_1 - {\bf p}_A ) \delta^3( {\bf k}_2 - {\bf p}_B ) }{V}
\bra{p_\gamma,p_\delta} i {\cal T} \ket{k_1, k_2}
\nonumber \\
&=&
\frac{1}{V \sqrt{s}} 
\,
(2 \pi)^4 \delta^4 ( \sum p_f - \sum \pin)
i {\cal M}^{\pin}_{p_f}\,,
\label{pfTpin}
\eea
where we used $s = 4 E_\alpha E_\beta$, which is valid at the centre of mass frame, and 
\be
\bra{p_\gamma,p_\delta} i {\cal T} \ket{k_1, k_2} = 
 (2 \pi)^4 \delta^4\big(\sum p_f - \sum \pin\big) \,
 i {\cal M}^{\pin}_{p_f}.
\label{S2M}
\ee
The $\varrho_x'$ in Eq.\ \eqref{QI2} is therefore calculated as
\be
\varrho_x' \,=\, \frac{T}{V} \frac{1}{s} \int_x d \Pi_{\rm LIPS} {\cal M}^{\pin}_{p_f} \left[ {\cal M}^{\pin}_{p_f} \right]^*
\,,
\label{varrho3}
\ee
where we introduced $VT = (2 \pi)^4 \delta^4(0)$, with $T$ being the interval over all time, 
and we have used $d \Pi_{\rm LIPS} =  (2 \pi)^4 \delta^4(\sum p_f - \sum \pin) d \Pi_{\gamma \delta}$.
This expression makes it clear that the outcome of the quantum instrument for collider measurements, $\varrho_x'$, vanishes in the limit of infinite volume and time-interval due to the dimension-full singular factor $T/V = 1/[(2 \pi)^2 \delta^2(0)]$.
It is also clear that this channel is trace non-increasing, i.e.\ $\Tr \varrho'_x < 1$.

To introduce a practically more useful non-singular quantity, $\varrho_x$, we cancel the $T/V$ factor and multiply it by $1/(2 \sigma_{\cal N})$ to compensate the dimension of $T/V$.
Namely, $\varrho_x$ in Eq.\ \eqref{varrho_form} is related with the direct quantum instrument outcome $\varrho_x'$ in Eq.\ \eqref{QI} as
\be
\varrho_x = \frac{V}{T} \frac{1}{2 \sigma_{\cal N}} \varrho_x' \,.
\ee
For convenience, we choose the normalisation factor $\sigma_{\cal N}$ to be the inclusive cross section for the maximally mixed initial state $\rin^{\rm mix}$, defined in Eq.\ \eqref{sigN}.
In this way, the scaling is independent of the initial state $\rin$ and the map ${\cal I}_x(\rin) = \varrho_x$ is still linear in $\rin$.

\subsection{Relation to the optical theorem}

In the previous subsection, we demonstrated that the outcome of the quantum instrument in collider physics, $\varrho_x'$, is proportional to the factor $T/V$, and its trace never reaches unity.
In this section, we clarify how the trace of the out-state is related to the probability conservation in particle physics, i.e.\ the unitarity of the $S$-matrix and the optical theorem.\footnote{See also \cite{Seki:2014cgq,Kowalska:2024kbs} for similar discussions.} 

As discussed in the step {\bf [3]} in section \ref{sec:QIcol}, the projection operator \eqref{proj} for the selective measurement is a part of the complete set defined in Eq.\ \eqref{compset}. 
Before the measurement, the out-state after scattering is $\tildepout = S \riSE S^\dagger$ in Eq.\ \eqref{preout}.
We now calculate $\Tr \tildepout$ using the complete set in \eqref{compset}.
We will show that the unitarity of the $S$-matrix implies $\Tr \tildepout = 1$.

As in the previous subsection, we concentrate on the momentum part.
Dropping the finite-dimensional spin/flavour part, $\Tr \tildepout$ is given by
\be
\Tr \tildepout =
\sum_f \left[ 
\left( \prod_{i \in f} \int d \Pi_i \right)
\bra{f} S | \tildepin \ketbra \tildepin | S^\dagger \ket{f}
\right]\,.
\ee
Using the completeness relation \eqref{compset}, the right-hand-side is equal to
$
\bra{\tildepin} S^\dagger S \ket{\tildepin}
$
and $\Tr \tildepout = \braket{ \tildepin | \tildepin} = 1$ follows if the $S$-matrix is unitary, $S^\dagger S = {\bf 1}$. 

For a different path, we split the sum of the final states into the state in which the initial particles and momenta are unchanged, i.e.\ $\ket{f} = \ket{p_\alpha,p_\beta}$ and the rest.
We also split the $S$ operator as $S = 1 + i {\cal T}$, where $1$ is the trivial non-interacting part and ${\cal T}$ is the interacting part, proportional to the coupling constants.  
The trivial part survives only for $\ket{f} = \ket{p_\alpha,p_\beta}$.
We have
\bea
\Tr \tildepout
&=& 
 \int d \Pi'_{\alpha \beta} 
\bra{p'_\alpha,p'_\beta} (1 + i {\cal T}) | \tildepin \ketbra \tildepin | (1 - i{\cal T}^\dagger) \ket{p_\alpha',p_\beta'}
\nonumber \\
&+& 
\sum_{f \neq \alpha \beta} \left[ 
\left( \prod_{i \in f} \int d \Pi_i \right)
\bra{f} {\cal T} | \tildepin \ketbra \tildepin | {\cal T}^\dagger \ket{f}
\right]
\eea

Using Eqs.\ \eqref{ptilde} and \eqref{pwave}, we obtain
\be
\braket{ p_\alpha',p_\beta' | \tildepin}
= \frac{\sqrt{s}}{V} (2 \pi)^6
\delta^3( {\bf p}_\alpha' - {\bf p}_\alpha )
\delta^3( {\bf p}_\beta' - {\bf p}_\beta )\,.
\ee
The zero-th order part in ${\cal T}$ of $\Tr \tildepout$ is therefore
\be
\Tr \tildepout \big|_{{\cal T}^0} =
\int d \Pi_{AB}' \left| 
\braket{ p_A',p_B' | \tildepin}
\right|^2 = 1\,.
\ee
This already saturates the appropriate magnitude of $\Tr \tildepout$. 

To compute the first order part in ${\cal T}$, we note
\be
\bra{p_\alpha, p_\beta} i {\cal T} \ket{\tildepin}
= \frac{T}{\sqrt{s}}  i {\cal M}^{\pin}_{\pin},
\ee
where ${\cal M }^{\pin}_{\pin}$
is the forward scattering amplitude involving interactions, and Eq.\ \eqref{S2M} and
$VT = (2 \pi)^4 \delta^4(0)$ were used.
At ${\cal T}^1$ order, we have
\be
\Tr \tildepout \big|_{{\cal T}^1} = -
\frac{T}{V} \frac{2}{s} {\rm Im} \left[ 
{\cal M }^{\pin}_{\pin}
\right]
\ee

The calculation for the second order part in ${\cal T}$ is analogous to that for Eqs.\ \eqref{pfTpin}, \eqref{S2M} and \eqref{varrho3}.
We obtain 
\bea
\Tr \tildepout \big|_{{\cal T}^2} &=& 
\frac{T}{V}
\sum_{f \neq (p_A,p_B)}
\frac{1}{s} \int d \Pi_{\rm LIPS} \left| {\cal M}^{\pin}_f \right|^2
\nonumber \\
&=&
\frac{T}{V} \cdot 2 \cdot \left[
\sum_{f \neq (p_A,p_B)}
\sigma(\pin \to f) \right] \,.
\eea
In the bracket, we have the inclusive cross section, $\pin \to {\rm anything}$, except for the forward scattering, i.e.\ $\ket{f} = \ket{p_\alpha,p_\beta}$. 

Collecting all the terms, we finally get
\be
\Tr \tildepout = 1 + \frac{T}{V} \cdot 2 \cdot 
\left[ 
\sum_{f \neq (p_A,p_B)}
\sigma(\pin \to f) 
-
\frac{1}{s} {\rm Im} \left[ 
{\cal M }^{\pin}_{\pin}
\right]
\right] \,.
\ee
In this expression, it is clear that $\Tr \tildepout = 1$ implies the optical theorem
\be
\sum_{f \neq (p_A,p_B)}
\sigma(\pin \to f) 
=
\frac{1}{s} {\rm Im} \left[ 
{\cal M }^{\pin}_{\pin}
\right] \,,
\ee
which, as it is well known, is the direct consequence of the unitarity of the $S$ operator, $S^\dagger S = 1 \Rightarrow {\cal T}^\dagger {\cal T} = i ({\cal T} - {\cal T}^\dagger)$.

\section{Numerical values for the coefficients $a_{ij}^{(*)}$ and $D_i$ appearing in Eqs.\ \eqref{Imatrix} and \eqref{traceChoi}}
\label{numerical}

We now report the values of $a_{ij}^{(*)}$ and $D_i$ coefficients appearing in Eqs.\ \eqref{Imatrix} and \eqref{traceChoi}, evaluated at tree-level in the Standard Model, outlined in Subsection \ref{sec:prediction},
and using measured values for the physical constants.  
Near the threshold region with $\sqrt{s} = 370$ GeV, we obtain
\bea
a^{(+)} \big|_{\sqrt{s} = 370\,{\rm GeV}} &=&
\begin{pmatrix}
0.503 & -1.442 & 0.364 & 0.503 \\
-1.442 & 4.137 & -1.043 & -1.442 \\
0.364 & -1.043 & 0.263 & 0.364 \\
0.503 & -1.442 & 0.364 & 0.503 
\end{pmatrix} \cdot 10^{-2} \,,
\nonumber \\
a^{(+-)}\big|_{\sqrt{s} = 370\,{\rm GeV}} &=&
\begin{pmatrix}
0.800 & 0.264 & -1.979 & 0.800 \\
-2.293 & -0.758 & 5.676 & -2.293 \\
0.578 & 0.191 & -1.431 & 0.578 \\
0.800 & 0.264 & -1.980 & 0.800 
\end{pmatrix} \cdot 10^{-2} \,,
\nonumber \\
a^{(-)}\big|_{\sqrt{s} = 370\,{\rm GeV}} &=&
\begin{pmatrix}
1.271 & 0.420 & -3.146 & 1.271 \\
0.420 & 0.139 & -1.040 & 0.420 \\
-3.146 & -1.040 & 7.788 & -3.146 \\
1.271 & 0.420 & -3.146 & 1.271 
\end{pmatrix} \cdot 10^{-2}\,,
\eea
\be
(D_0, D_1, D_2)|_{\sqrt{s} = 370\,{\rm GeV}} ~=~ 
(0.159,\, 0.115, \, 0.0878 )\,.
\nonumber
\ee
while for $\sqrt{s} = 1$ TeV, these values change to 
\bea
a^{(+)} \big|_{\sqrt{s} = 1\,{\rm TeV}} &=&
\begin{pmatrix}
0.219 & -0.777 & -0.494 & 0.219 \\
-0.777 & 2.755 & 1.751 & -0.777 \\
-0.494 & 1.751 & 1.113 & -0.494 \\
0.219 & -0.777 & -0.494 & 0.219
\end{pmatrix} \cdot 10^{-2} \,,
\nonumber \\
a^{(+-)}\big|_{\sqrt{s} = 1\,{\rm TeV}} &=&
\begin{pmatrix}
0.340 & -0.810 & -1.162 & 0.340 \\
-1.205 & 2.870 & 4.117 & -1.205 \\
-0.766 & 1.824 & 2.617 & -0.766 \\
0.340 & -0.810 & -1.162 & 0.340 
\end{pmatrix} \cdot 10^{-2} \,,
\nonumber \\
a^{(-)}\big|_{\sqrt{s} = 1\,{\rm TeV}} &=&
\begin{pmatrix}
0.527 & -1.256 & -1.802 & 0.527 \\
-1.256 & 2.990 & 4.290 & -1.256 \\
-1.802 & 4.290 & 6.154 & -1.802 \\
0.527 & -1.256 & -1.802 & 0.527 
\end{pmatrix}\cdot 10^{-2} \,,
\eea
\be
(D_0,D_1,D_2) |_{\sqrt{s} = 1\,{\rm TeV}} \,=\, 
(0.145, \,  0.0481, \, 0.115)\,.
\nonumber
\ee

\bibliographystyle{JHEP}
\bibliography{refs}

\end{document}